\documentclass[a4paper,fleqn,usenatbib]{mnras}

\usepackage[T1]{fontenc}
\usepackage{ae,aecompl}

\usepackage{graphicx}	% Including figure files
\usepackage{amsmath}	% Advanced maths commands
\usepackage{amssymb}	% Extra maths symbols
\usepackage{subfig}		% Sub figures
\usepackage[title]{appendix}
\usepackage{cleveref}	% 
\usepackage{geometry}
\usepackage{siunitx}
\usepackage{lipsum}

\newcommand{\vecb}[1]{\boldsymbol{#1}}	% Vectors
\newcommand{\avg}[1]{\left\langle #1 \right\rangle}
\renewcommand{\tilde}{\widetilde}

\graphicspath{{./}{figures/}}
\crefname{figure}{Fig.}{Figs.}
\Crefname{figure}{Fig.}{Figs.}
\crefname{appsec}{Appendix}{Appendices}

\title[Quantifying EoR delay spectrum contamination]{Quantifying EoR delay spectrum contamination from diffuse radio emission}

\author[A. E. Lanman et al.]{
Adam E. Lanman,$^{1}$\thanks{E-mail: adam\_lanman@brown.edu}
Jonathan C. Pober,$^{1}$
Nicholas S. Kern,$^{2}$
Eloy de Lera Acedo,$^{3}$
\newauthor
David R. DeBoer,$^{2}$
Nicolas Fagnoni$^{3}$
\\
$^{1}$ Department of Physics, Brown University, Providence, RI\\
$^{2}$ Department of Astronomy, University of California, Berkeley, CA\\
$^{3}$ Cavendish Astrophysics, University of Cambridge, Cambridge, UK
}

\pubyear{2019}

\begin{document}
\label{firstpage}
\pagerange{\pageref{firstpage}--\pageref{lastpage}}
\maketitle

\begin{abstract}
The 21~cm hyperfine transition of neutral hydrogen offers a promising probe of the large scale structure of the universe before and during the Epoch of Reionization, when the first ionizing sources formed. Bright radio emission from foreground sources remains the biggest obstacle to detecting the faint 21~cm signal. However, the expected smoothness of foreground power leaves a clean window in Fourier space where the EoR signal can potentially be seen over thermal noise. Though the boundary of this window is well-defined in principle, spectral structure in foreground sources, instrumental chromaticity, and choice of spectral weighting in analysis all affect how much foreground power spills over into the EoR window. In this paper, we run a suite of numerical simulations of wide-field visibility measurements, with a variety of diffuse foreground models and instrument configurations, and measure the extent of contaminated Fourier modes in the EoR window using a delay-transform approach to estimating power spectra. We also test these effects with a model of the HERA antenna beam generated from electromagnetic simulations, to take into account further chromatic effects in the real instrument. We find that foreground power spillover is dominated by the so-called ``pitchfork effect'', in which diffuse foreground power is brightened near the horizon due to the shortening of baselines. As a result, the extent of contaminated modes in the EoR window is largely constant over time, except when the galaxy is near the pointing center.
\end{abstract}

\begin{keywords}
techniques: interferometric -- dark ages, reionization, first stars, instrumentation: interferometers, methods: numerical
\end{keywords}

%%%%%%%%%%%%%%%%%%%%%%%%%%%%%%%%%%%%%%%%%%%%%%%%%%

%%%%%%%%%%%%%%%%% BODY OF PAPER %%%%%%%%%%%%%%%%%%

\section{Introduction}

The Epoch of Reionization (EoR) comprises the period of the Universe's history after recombination when the first luminous structures formed, emitting UV radiation that carved out ionized regions of the neutral intergalactic medium (IGM). The lack of bright sources and optically-thick IGM during the the EoR limit observational prospects through traditional high-redshift galaxy surveys and quasar absorption measurements. The hyperfine transition of neutral hydrogen, which emits a rest-frame photon with a wavelength of 21 cm, offers a promising probe of the IGM structure during the EoR and before, when neutral hydrogen was abundant in the Universe (see \cite{pritchard_evolution_2008, furlanetto_cosmology_2006} for a review). As a forbidden transition, the IGM is largely transparent to 21 cm photons, and so this signal is not obscured by intervening gas. As a line transition, the redshift of 21 cm emission/absorption directly corresponds to distance. In this way, mapping the distribution of redshifted 21 cm brightness can map the three-dimensional structure of neutral hydrogen during the EoR.

%Weak signal -- Imaging is difficult, but power spectrum measurements are more feasible.
Directly imaging structures of the EoR will likely require an instrument of the scale of the upcoming Square Kilometre Array (SKA; \citealt{mellema_reionization_2013, furlanetto_cosmology_2006}), so current generation experiments are seeking statistical measures of the EoR signal, such as the power spectrum. Radio interferometers are well suited to power spectrum measurements, since they directly sample the angular Fourier transform of the sky, and can achieve high spectral resolution, which provides good sampling of line-of-sight Fourier modes. Many of these instrument feature highly redundant, compact layouts that maximize sensitivity to large angular scales, aiming to detect the EoR signal through their fine spectral resolution first \citep{parsons_sensitivity_2012}. Though an actual detection of the power spectrum has yet to be made, experiments like the GMRT \citep{paciga_simulation-calibrated_2013, paciga_gmrt_2011}, MWA \citep{barry_improving_2019, beardsley_first_2016, bowman_science_2013}, PAPER \citep{kolopanis_simplified_2019,cheng_characterizing_2018, parsons_precision_2010}, and LOFAR \citep{patil_upper_2017}, have placed upper limits on the power spectrum amplitude at several redshifts.

%Foreground power is a considerable hurdle to detecting the EoR. (forms it takes, behavior.)
The primary obstacle to these experiments is contamination of bright foreground sources, such as Galactic synchrotron emission, free-free emission, radio galaxies, pulsars, and supernova remnants, which are typically five orders of magnitude brighter than the expected EoR signal \citep{furlanetto_cosmology_2006,  oh_foregrounds_2003, matteo_radio_2002}. Fortunately, the smooth spectra exhibited by these sources distinguish them from the relatively complex structure of EoR signals, making it possible -- in principle -- to model and subtract them from data \citep{oh_foregrounds_2003, liu_how_2012, chapman_cosmic_2015, sims_contamination_2016, carroll_high_2016} or avoid them in Fourier space \citep{parsons_per-baseline_2012, parsons_sensitivity_2012, pober_opening_2013, chapman_effect_2016}. Since frequency maps directly to redshift and distance, this separation occurs in Fourier space parallel to the line of sight ($k_\parallel$). Smooth-spectrum power will cluster around $k_\parallel = 0$, while EoR signal power will spread farther.

Foreground mitigation strategies that rely on the spectral smoothness of foreground sources must also account for instrumental sources of spectral structure. One such effect is due to the natural chromaticity of an interferometer caused by the effective change of baseline length with frequency. A baseline samples angular scales on a scale inversely proportional to the baseline length in wavelengths, which means that a baseline will sample finer scales at higher frequencies. This is equivalent to the baseline drifting through Fourier space with frequency, which creates a frequency dependence in the visibility measurements. This couples $k_\perp$ and $k_\parallel$ and spreads power to higher $k_\parallel$ modes \citep{datta_bright_2010, vedantham_imaging_2012, morales_four_2012, parsons_per-baseline_2012, trott_impact_2012, thyagarajan_study_2013, liu_epoch_2014, pober_what_2014}.

Though baseline chromaticity itself is fundamental to the measurement, its effects are confined to a well-defined region of Fourier space called the \emph{foreground wedge}, with a maximum $k_\parallel$ set by the angular extent of the field of view, called the ``horizon limit.'' This leaves a region called the \emph{EoR window} which should be uncontaminated by foregrounds. Nonetheless, other sources of spectral structure, including the intrinsic spectra of foreground sources, the finite bandpass of the instrument, and the chromaticity of the antenna gains can spread foreground power beyond the wedge into the window. This suprahorizon power is usually accounted for in foreground avoidance approaches by avoiding $k_\parallel$ modes within some buffer distance of the horizon line.

Estimates of an appropriate buffer distance have been largely motivated by simple, conservative  models and observations in data. \cite{parsons_per-baseline_2012} argued that foreground power spillover in delay spectra is caused by the convolution of the spectral responses of the beam and sky brightness, and so is limited by the inverse bandwidth of the instrument. Through simplified foreground simulations they demonstrate that this spillover does not exceed $k_\parallel \sim 0.13$~Mpc$^{-1}$. \cite{thyagarajan_study_2013} examined the effect of spectral windowing on this spillover using numerical formulations of point source and noise power spectra, finding that increased foreground suppression came at the expense of narrowing the ``effective'' bandwidth, and hence increasing the extent to which foreground power spills into the EoR window. \cite{pober_opening_2013} observed suprahorizon spillover in power spectra taken from PAPER data, finding that the spillover is not constant with $k_\perp$ (baseline length), and is larger on short baselines. Imaging the data after applying a high-pass filter to remove the wedge, they found that remaining contamination is dominated by unresolved diffuse power. 

More recently, \cite{thyagarajan_effects_2016} used wide-field visibility simulations with delay spectrum analysis to measure the levels of foreground power in the EoR window for baselines in the 19-element deployment of HERA. These simulations used a HERA beam model made through electromagnetic simulation of a HERA dish and receiver element, which incorporates spectral structure due to reflections among the various components of the antenna. They found that, except when the galaxy is directly overhead, foreground power can be suppressed below the EoR amplitude for $k_\parallel \geq 0.2 $~h~Mpc$^{-1}$.

Though these earlier works predicted that foreground power should extend  as far as $\Delta k_\parallel \sim 0.2 $~h~Mpc$^{-1}$, published power spectrum limits usually choose narrower buffers when binning measurements in $k$. \cite{dillon_empirical_2015} used a moderate buffer of $\Delta k_\parallel \sim 0.02 $~h~Mpc$^{-1}$ when binning MWA power spectra in 1D, which noticeably left in some suprahorizon emission. \cite{ali_paper-64_2015} applied a buffer of 15~ns when binning power spectra for PAPER-64 data, corresponding with $\Delta k_\parallel \sim 0.008$~h~Mpc$^{-1}$ at their redshift  of $z \sim 8.74$. This buffer was set by the inverse of the bandwidth used for the analysis, and did not take into account any expectations of spectral structure in the instrument or sky. Both \cite{beardsley_first_2016} and the more recent \cite{barry_improving_2019} account for suprahorizon emission by increasing the slope of the horizon line by 14\%, the horizon line being the line in $(k_\perp, k_\parallel)$ space corresponding with the horizon limit. This choice makes for a wider buffer at large $k_\perp$, which runs contrary to the expectation that foreground spillover is larger for small $k_\perp$.

% Note -- converting their results to non-little-h units using Planck15 little-h

% 		Thyagarajan 2015 -- Describe a "strip" near the horizon with width inversely proportional to the bandwidth. (thyagarajan_foregrounds_2015)
%		Thyagarajan 2013 -- Examines how blackman-nutall and extended BN window compare with rectangular.  (thyagarajan_study_2013)
% 		Pober 2015  "What next generation..." (pober_what_2014) -- Sets up a background on the topic.
% Parsons 2012 ("A per baseline...") Fig 8 shows the simulated pspecs of foregrounds. Argues that k ~ 0.2 h Mpc^-1 is the upper limit of fg contamination for baselines shorter than 16λ  (parsons_per-baseline_2012)
% 		Pober 2014 "The impact of foreground..." (pober_impact_2015)  and Pober 2013b cited therein ("Opening the 21cm EoR window...") (pober_opening_2013) --- sees suprahorizon emission in PAPER data.

% 		Dillon 2015 "Mapmaking for..." (dillon_mapmaking_2015) --- Calls it a buffer, cites Pober 2013 for it.
% 		Dillon 2015 "Empirical Covariance Modeling..." --- cites the 0.1 to 0.15 h Mpc^-1 buffer, chooses a conservative 0.5. (dillon_empirical_2015)
% 		Ali Z. S.et al. 2015 ApJ  809 61 -- 15 ns buffer, cites a couple sources on that (PAPER 64 data) (ali_paper-64_2015) --- Also look at Matt's revised PAPER-64 results. What did he use?
% 		Beardsley 2016 ("First season...") (beardsley_first_2016) -- Buffer set by increasing slope of horizon line by 14%. (Nichole and Wenyang?)
%           Barry 2019 uses Dillon's 14% slope increase buffer. (barry_improving_2019)

Avoiding foreground-contaminated modes is essential to making a robust EoR power spectrum measurement, so it is advisable to set a wide buffer beyond the horizon limit. However, setting too wide a buffer means potentially throwing out the most sensitive measurements. It is therefore essential to have precise predictions of how foreground power spreads into the EoR window for different instruments in order to use foreground avoidance techniques.

In this paper, we simulate visibility measurements from a zenith-pointing array with realistic, frequency-dependent antenna beams, observing a variety of diffuse foreground models. From these simulated observations we make power spectra using delay-transform methods and measure how far beyond the horizon foreground power spills over as a result of a the effects included in simulations. Such effects include the spectral structure of the primary beam or foreground model, baseline length, primary beam shape, and proximity of the brightest emission to the beam center. We focus on diffuse foreground models because EoR power spectrum estimations typically only use the shortest baselines in the array, which are both most sensitive to the expected EoR signal and to diffuse foregrounds.

The following section gives some background on the 21~cm power spectrum, the delay transform estimation approach, and the sources of foreground power spillover. We also discuss the theoretical explanation for the foreground pitchfork, and discuss the choice of simulations to explore these various effects. \Cref{sec:simulator} discusses the numerical simulator and the instrument and sky model configurations tested. \Cref{subsec:pitchfork_conf,subsec:window_func_comp,subsec:spec_index_comp,subsec:bleed_results} presents measurements of the effects of baseline foreshortening near the horizon, spectral windowing, source and beam spectral structure, and source position and baseline length. Lastly, we show the same results with a more realistic primary beam model in \cref{subsec:cst_beam_results}.

\section{Diffuse Foreground Power}
\label{sec:21cm_pspec_general}

The 21 cm power spectrum $P_{21}$ is defined as
\begin{equation}
\avg{\tilde{T}_{b} (\vecb{k}) \tilde{T}^*_{b} (\vecb{k}') }
= (2 \pi)^3 \delta_D(\vecb{k} - \vecb{k}') P_{21}(\vecb{k})
\end{equation}
where $T_b$ is the brightness temperature contrast of the 21~cm line against the background radiation field, tildes denote the spatial Fourier transform and $\delta_D$ is the three-dimensional Dirac delta function. The wave vector $\vecb{k}$ may be broken down into $\vecb{k}_\perp = (k_x, k_y)$, which is perpendicular to the line of sight, and $k_\parallel$ which is parallel to the line of sight. We can form an estimator of the power spectrum from a measured volume $\mathbb{V}$.
\begin{equation}
\hat{P}(\vecb{k}) \equiv \frac{\avg{|\tilde{T}(\vecb{k})|^2}}{\mathbb{V}}
\label{eqn:general_pspec}
\end{equation}

%Though the true 21~cm signal is isotropic, power spectrum measurements are often expressed in a cylindrically-binned basis, in terms of $k_\perp = |\vecb{k}_\perp|$. Each baseline in an interferometers effectively probes a single $k_\perp$ mode, while sampling the $k_\parallel$ axis through its spectral response,w

\subsection{Delay Transform}

There are a variety of methods available for estimating $\tilde{T}(\vecb{k})$ from interferometric measurements, but they can be very generally categorized as either based on the delay transform on sky-reconstruction (see \citealt{morales_understanding_2019} for a summary). In both categories, $k_\parallel$ modes are accessed by Fourier-transforming over an axis probed by frequency. We will focus here on the delay spectrum approach \citep{parsons_per-baseline_2012, parsons_sensitivity_2012, parsons_new_2014}.

The complex visibility measured by a single baseline is given by the radio interferometer measurement equation (RIME)  \citep{thompson_interferometry_2017},
\begin{equation}
V(\vecb{b}, \nu) = \int\limits_\text{sky} A(\hat{s},\nu) T_b(\hat{s},\nu) e^{-2 \pi i ((\hat{s} - \hat{s}_p) \cdot \vecb{b}) \nu/c} d^2 s,
\label{eqn:rime_full}
\end{equation}
where $\nu$ is the frequency, $c$ is the speed of light, $\vecb{b}$ is the baseline vector, and $A$ is the primary beam function giving directional gains of the measuring antennas. $\hat{s}_p$ is a unit vector pointing to the \emph{phase center}, and the integral is carried out over all unit vectors $\hat{s}$ over half of the unit sphere, denoted by ``sky''. The phase center term cancels out of the eventual estimator, so we omit it going forward.

The delay transform is an inverse Fourier transform applied along the frequency axis: 
\begin{equation}
\widetilde{V}(\vecb{b}, \tau) = \int\limits_{-\infty}^\infty \int\limits_\text{sky} \phi(\nu) A(\hat{s},\nu) T(\hat{s},\nu)
 e^{-2 \pi i (\tau_g - \tau) \nu} d^2s\: d\nu
\label{eqn:dt1}
\end{equation}
The function $\phi(\nu)$ is a spectral window, which both enforces the finiteness of the bandpass and allows us to applying weighting to each frequency. Here $\tau_g = \hat{s} \cdot \vecb{b} \nu/c$ is the \emph{geometric delay}, corresponding to the time it takes a wavefront propagating from position $\hat{s}$ to cross the baseline. Under a flat-sky approximation, the integral over the sky is equivalent to a 2D Fourier transform from $\hat{s}$ to $\vecb{u}$, sampled at the baseline position $\vecb{u} = \vecb{b} \nu/c$. The delay $\tau$, dual to frequency, is then an approximate probe of line of sight Fourier modes.\footnote{This correspondence is only approximate because the baseline length in wavelengths changes with frequency. Hence, the delay modes are not perfectly orthogonal to the $k_\perp$ modes.} The delay transformed visibility is thus related to the Fourier-transformed temperature $\tilde{T}_b$, convolved with the beam and taper function.

With appropriate cosmological scaling, we can form a power spectrum estimator.
\begin{equation}
\hat{P}(\vecb{k}_\perp, k_\parallel) \equiv \frac{X^2 Y}{B_{pp} \Omega_{pp}} |\tilde{V}(\tau)|^2
\label{eqn:pspec_est}
\end{equation}
The multiplicative factors in front approximate the inverse volume factor of \cref{eqn:general_pspec}. $X$ and $Y$ are redshift-dependent cosmological scaling factors, with units of length per angle and length per frequency, respectively. $B_{pp}$ is an effective bandwidth, given by $B_{pp} = \int |\phi(\nu)|^2 d\nu$, which relates to the total line of sight extent observed. The primary beam squared integral $\Omega_{pp} = \int |A(\hat{s})|^2$ gives the angular area. The exact derivation of this factor is given by Appendix B of \citep{parsons_new_2014}.\footnote{See also Memos \#27 and \#43 at reionization.org.}

The cosmological factors are
\begin{align}
X(z) = \chi(z) \qquad Y(z) = \frac{c (1+z)^2}{H(z) \nu_{21}},
\end{align}
where $z$ is redshift, $\chi$ is the comoving distance, $H(z)$ is the Hubble parameter, $\nu_{21}$ is the rest-frame 21 cm frequency, and $c$ is the speed of light. The $\vecb{k}_\perp$ and $k_\parallel$ modes corresponding to a given baseline $\vecb{u}$ and delay $\tau$ are given by
\begin{align}
\vecb{k}_\perp = \frac{2\pi \vecb{u}}{X} \qquad k_\parallel &=  \frac{2 \pi \tau}{Y}
\end{align}
For the above relations, $X$ and $Y$ are evaluated at the center of the bandpass and assumed not to evolve much over the chosen bandpass. As mentioned before, the correspondence between $k_\parallel$ and $\tau$ is only approximate, and gets worse for longer baselines. Throughout this paper, we will use a $\Lambda$CDM cosmological model with parameters measured by \cite{planck_2015}.

\subsection{Foreground Power Spillover}
\label{subsec:fg_power_spillover}
In the case of a single point source at position $\hat{s}_p$, with corresponding geometric delay $\tau_p$, and spectrum $I_p(\nu)$, \cref{eqn:dt1} reduces to
\begin{equation}
\widetilde{V}(\vecb{b}, \tau) = (\widetilde{\phi} \ast \widetilde{A} \ast \widetilde{I}_p)(\tau - \tau_p)
\label{eqn:dt2}
\end{equation}
where $\ast$ denotes convolution in delay and tildes indicate a delay-transformed function. If the beam, source spectra, and window function are all flat in frequency, and the bandpass is infinite, then this becomes a delta function centered at the geometric delay of the point source. The maximum geometric delay is limited to the baseline length $b = |\vecb{b}|$ over the speed of light.
\begin{equation}
|\tau_g| \leq \frac{b}{c}
\end{equation}
Under these conditions, the power from a point source on the horizon will has the form of a delta function centered at this maximum delay. Hence, this maximum delay is known as the ``horizon line'', and separates flat-spectrum power from spectrally-structured emission. In cosmological terms, this horizon limit is given by
\begin{equation}
k_\parallel \leq \frac{H(z) \chi(z)}{c (1+z)} k_\perp
\end{equation}
where $k_\perp = |\vecb{k}_\perp|$.

This limit applies in the ideal point source case as discussed --- infinite bandwidth and a flat spectrum. This power will spread out, possibly beyond the horizon limit, once these assumptions are relaxed. For example, assume the primary beam $A$ and point source flux $I_p$ have no spectral structure, but the window function has a limited domain. In this case, $\widetilde{A}(\tau) = A_p \delta_D(\tau)$ and $\widetilde{I}_p(\tau) = I_p \delta_D(\tau)$, where $\delta_D$ is the Dirac delta function, and \cref{eqn:dt2} reduces to 
\begin{equation}
\widetilde{V}(\vecb{b}, \tau) = I_p A_p \widetilde{\phi}(\tau - \tau_p)
\end{equation}
If $\phi$ is a rectangular window function, then $\tilde{\phi}$ is a sinc function centered on $\tau = \tau_p$, which has sidelobes that are not confined to the horizon. The sinc function is known to lack the dynamic range needed to access the expected EoR amplitude, so most analyses use other common window functions such as Hamming, Blackman-Harris, or Blackman-Nuttall. These windows suppress their sidelobe amplitude at the expense of widening the main lobe. In any case, the finiteness of the bandpass necessarily causes even flat-spectrum power to spread.

%\Cref{fig:sincleakage_diag} shows the delay spectrum for a single flat-spectrum point source on the horizon with a finite rectangular bandpass. The sidelobes of the sinc function reach non-negligible amplitude beyond the horizon limit.

% \begin{figure}
%     \centering
%     \includegraphics[width=0.45\textwidth]{sinc_leakage_diagram.eps}
%     \caption{The spread of power power beyond the horizon limit (black line) for a flat spectrum point source on the horizon with a finite bandpass (orange), along with the power spectrum of a noise-like EoR signal (blue) of appropriate amplitude. The sidelobes of the sinc function can overwhelm EoR signals beyond the horizon line (to the right).}
%     \label{fig:sincleakage_diag}
% \end{figure}

\subsection{Foreground Pitchfork}
\label{sec:pitchfork}

Resolved diffuse emission behaves in a fundamentally different way from unresolved point-like sources in interferometric measurements. The difference can be illustrated in angular Fourier space by taking the flat sky approximation of \cref{eqn:rime_full} such that the exponential fringe term acts as a 2D Fourier transform. Applying the convolution theorem, we can see that the visibility measured by a baseline is given by the convolution of the primary beam and sky brightness in Fourier space:
\begin{align}
V(u,v) &\simeq \int A(l,m) T(l,m) e^{-2 \pi i (ul + vm)} dl dm\\
	&= (\widetilde{A} \ast \widetilde{T}) (u,v)
\end{align}
Here, $(l,m)$ are direction cosines, tildes denote the 2D Fourier transform and $(u,v)$ are orthogonal components of the baseline vector measured in wavelengths. These are also the Fourier-duals to $l$ and $m$. A point source has a surface brightness $I$ in the form of a delta function, which is constant in Fourier space. A diffuse source, however, will be more compact in Fourier space, centered on $u,v \approx 0$. Short baselines will therefore measure more diffuse power than long baselines, while the measured power from point sources will be the same for both.

The increased brightness of diffuse power on short baselines leads to an increase in measured brightness when diffuse sources are near the horizon, due to the shorter projected baseline length. This puts excess power near the horizon in delay space, leading to a characteristic pitchfork pattern, observed in simulations and in MWA data \citep{thyagarajan_foregrounds_2015, thyagarajan_confirmation_2015}. Since the projected baseline length can go to effectively zero at the horizon, even a monopole signal, normally undetectable to interferometers, produces a signal.

The pitchfork effect can potentially change our intuition for how foreground power spreads beyond the horizon. For a zenith-tracking array, measured power is highest when bright sources are near zenith, because the primary antenna beam is strongest there. In delay space, however, this power is clustered around $\tau=0$, so how much of that power spreads beyond the horizon depends on the spectral structure of the emission and the instrument. Power in the pitchfork is very close to the horizon, but is attenuated by the weaker primary beam there. This raises a question -- Does a bright patch of the sky contaminate the window more when it's at zenith or near the horizon?

To answer this, we will look at how much power spreads beyond the horizon as a function of the hour angle of the galactic center, for a variety of primary beam models, in simulated data. Some of these beam models have effectively zero response near the horizon, which prevents any pitchfork from appearing. Comparing these, we can demonstrate that the pitchfork is the dominant source of power near the horizon for most observing time, while spectral structure in the beam or source affects how that power is spread. By taking both effects into consideration, it is possible to set bounds on the extent of foreground contamination.

\section{Simulations}
\label{sec:simulator}

\begin{table*}
\begin{tabular}{|c|p{4in}|}
\hline
\multicolumn{1}{|l}{\textbf{Beam Models}} &  \\ \hline
Airy Disk & Radiation pattern of a circular aperture with a 14 m diameter. \cref{eqn:airy} \\ \hline
Achromatic Gaussian & $\exp(-\theta^2/(2 \sigma_0^2))$, with $\sigma_0 = 7.37^\circ$. This is set to match the FWHM of the main lobe of the Airy beam. \cref{eqn:agauss} \\ \hline
Chromatic Gaussian & The same as achromatic, but with $\sigma  = \sigma_0 (\nu/\nu_0)^\alpha$ \cref{eqn:cgauss} \\ \hline
HERA CST & Power beams made from far-field electric fields measured in CST simulations of a HERA dish with crossed-dipole feeds. \\ \hline
\multicolumn{1}{|l}{\textbf{Sky Models}} &  \\ \hline
GSM & The Global Sky Model in HEALPix, generated from three principal components per pixel \citep{zheng_improved_2017}. \\ \hline
%aGSM & An achromatic GSM. The value of each pixel is a constant across the bandpass, set as the mean of the pixel's spectrum in the chromatic GSM. \\ \hline
sym & A model with the same angular power spectrum as the GSM, defined to be azimuthally symmetric about the galactic center. Each pixel is given a power law frequency spectrum.\\ \hline
\multicolumn{1}{|l}{\textbf{Layout}} &  \\ \hline
37 hex & A thirty-seven element hexagon, with 14.6 m spacing. \\ \hline
Imaging & 80 antennas distributed randomly, with a Gaussian radial density profile with FWHM of 230 m. The longest baseline is 608 m.\\ \hline
Line & A line of 300 antennas, evenly spaced, east to west. Baselines are formed by pairing each of these antennas with a 301st antenna, such that the whole array provides East/West baselines with lengths ranging from 10 to 150 m. \\ \hline
\multicolumn{1}{|l}{\textbf{Time}} &  \\ \hline
Off-Zenith & Five 11s timesteps when the Galactic center is 4 hours off of zenith. \\ \hline
Transit & 24 hours at 5 minute timestep, such that the galactic center transits 6 hours from the start. \\ \hline
\multicolumn{1}{|l}{\textbf{Frequency}} &  \\ \hline
50 MHz & 100 to 150 MHz with channels of width 97.65 kHz. The channel width is chosen to match that of the first generations of HERA. \\ \hline
\end{tabular}
\caption{A summary of parameters used in simulations.}
\label{tab:sim_params}
\end{table*}

Simulations were carried out by numerically integrating the radio interferometry measurement equation (RIME). The integration is done from horizon to horizon, taking into account the curvature of the sky. The simulator code is publicly available \citep{lanman_healvis:2019}.\footnote{https://github.com/RadioAstronomySoftwareGroup/healvis} The brightness temperature $T(\hat{s}$ describes the specific intensity as a function of angle. We can discretize the sky by dividing it into pixels $p$, each of which is defined by a function $W_p(\hat{s})$ which is $1$ when $\hat{s}$ points into the pixel and $0$ otherwise. The discretized form of the RIME (\cref{eqn:rime_full}) has the form:
\begin{align}
V(\vecb{b}, \nu) &= \sum\limits_p \int\limits_\text{sky} W_p(\hat{s}) T(\hat{s}, \nu) A(\hat{s})
e^{-2\pi i \hat{s} \cdot \vecb{b} \nu/c} d\Omega \label{eqn:sim_int_0}\\
&\approx \sum\limits_p T(\hat{s}_p, \nu) A(\hat{s}_p)
e^{-2\pi i \hat{s}_p \cdot \vecb{b} \nu/c} \int\limits_\text{sky} W_p(\hat{s}) d\Omega
\label{eqn:sim_int_1}
\end{align}
The second step assumes that the fringe, surface brightness, and primary beam are all constant across the pixel, and so they can be moved outside the integral. The integral over each pixel window function then reduces to the solid angle of that pixel, $\omega_p$.

\begin{equation}
V_b(\nu) = \sum\limits_{p} \omega_p T(\hat{s}_p, \nu) A(\hat{s}_p)
e^{-2\pi i \hat{s}_p \cdot \vecb{b} \nu/c}
\label{eqn:sim_int}
\end{equation}

\Cref{eqn:sim_int} treats the diffuse sky model as a set of point sources at pixel centers with specific flux densities $T_p \omega_p$. We thus refer to this as the \emph{point source approximation}. As discussed in \cref{sec:pitchfork}, the difference in behavior between point sources and diffuse models is most significant on long baselines, so we need to ensure the map resolution is fine enough for this approximation to be valid. \Cref{app:appendix_point_source} discusses the limitations of the point source approximation and tests its validity for the simulations discussed here.

\Cref{tab:sim_params} summarizes the various simulation configurations used, including antenna layouts, primary beam models, sky brightness distribution models, observing time ranges and time step sizes. The same bandpass and channelization, covering 100 to 150~MHz at 97~kHz channel resolution, is used for all simulations. Generally, the ``Off-Zenith'' time configuration is used whenever LST is not considered important to the result. This time set consists of only five 11~s integrations, from a time when the Galactic center is in the sky but not near the beam center (off by 4 hours). This is considered as a typical sky brightness, an assumption supported by the tests with time dependence (using the ``transit'' time set). More details on the sky and beam models are given in the subsequent subsections.

\subsection{Primary Beam Models}

The primary beam $A(\hat{s})$ gives the direction-dependent gains of the antenna receiver elements, normalized to the value at the antenna's pointing center. The primary beam of a given antenna can be found by considering the antenna as a transmitter, fed by a uniform voltage, and measuring the far-field electric field pattern, and normalizing to the value at zenith. The \emph{power beam} is obtained by taking the squared amplitude of this electric field.

We use three analytically-defined beam models and one numerically calculated through electromagnetic simulations of a HERA dish. The analytic models are described in \crefrange{eqn:airy}{eqn:cgauss}.

The \emph{Airy} beam is defined as the far-field radiation pattern of a uniformly-illuminated circular aperture:
\begin{equation}
A_\text{airy}(\theta, \nu) = \frac{\nu/c}{\pi D \sin{\theta}}\: J_1 \left(\frac{\pi D \sin{\theta}}{\nu/c} \right)
\label{eqn:airy},
\end{equation}
where $\theta$ is the angle from the pointing center, $J_1$ is the first-order Bessel function of the first kind, $c$ is the speed of light, and $D$=14 m is the antenna aperture. The Airy beam has a simple analytic form that behaves similarly to a physical antenna, in that it has a non-negligible response near the horizon and has nulls that move with frequency.

The \emph{achromatic Gaussian} beam is defined by a Gaussian function with a constant full-width at half maximum (FWHM) across all frequencies, defined by a width parameter $\sigma_0$:

\begin{equation}
A_\text{AG} (\theta) = \exp \left(-\frac{\theta^2}{2\sigma_0^2}\right)
\label{eqn:agauss}
\end{equation}
This beam has no sidelobes and is effectively zero at the horizon. We choose $\sigma_0 = 7.37^\circ$, which sets the FWHM equal to the width of the main lobe of the Airy disk beam.

The chromatic Gaussian beam is defined just as \cref{eqn:agauss}, but with a width that evolves according to a power law.
\begin{align}
A_\text{CG}(\theta, \nu) &= \exp \left(-\frac{\theta^2}{2\sigma(\nu)^2}\right)\\
\sigma(\nu) &= \sigma_0 \left( \frac{\nu}{\nu_0}\right)^\alpha
\label{eqn:cgauss}
\end{align}
The reference frequency $\nu_0$ is chosen to be the first frequency channel center. Unless otherwise noted, the spectral index $s$ is chosen to be $-1$, which resembles the frequency evolution of the Airy beam.

The last beam model is defined numerically by a set of CST Suite\footnote{https://www.cst.com}  simulations of a realistic HERA dish and receiver done by \cite{fagnoni_hydrogen_2016}. The simulated antenna included a parabolic reflecting dish with a 14~m diameter, two 1.3~m copper dipole receivers with a pair of aluminum discs acting as sleeves, and a cylindrical mesh ``skirt'' 36~cm high around the feed to protect from reflections of other antennas in the array (see \cite{fagnoni_hydrogen_2016} for more details). Reflections among the various elements of the antenna introduce spectral structure into the beam, which is largely captured by the CST simulations.

These simulated beam models are stored as a set of electric field components at altitude and azimuth positions on a spherical grid with steps of 1$^\circ$ in both latitude and longitude, at frequencies between 50 and 250 MHz in steps of 1 MHz. For use in simulations of unpolarized (i.e., purely stokes-I) sky models, we convert this electric field beam into a \emph{power beam} in the following way. The electric field beam of an antenna feed $p$ may be expressed in a basis of unit vectors $\hat{\theta}$ and $\hat{\varphi}$ along the zenith angle and azimuthal directions, respectively:
\begin{equation}
\vecb{A}^p(\hat{s},\nu) = A^p_\theta(\hat{s}, \nu) \hat{\theta} + A^p_\varphi(\hat{s}, \nu) \hat{\varphi}
\end{equation}
The components form the elements of a Jones matrix, describing the directional response of the antenna feed to the electric field components of the sky. This is described in detail in \cite{kohn_hera-19_2018}. For unpolarized sky models, the response of each feed is given by the power of the electric field:
\begin{equation}
A^{pp}(\hat{s},\nu) = |A^p_\theta(\hat{s}, \nu)|^2 + |A^p_\varphi(\hat{s}, \nu)|^2
\label{eqn:uvbeam_power}
\end{equation}

The CST simulated beams have two crossed-dipole feeds, one oriented East/West (designated X) and the other oriented North/South (Y). Under the same formalism, we may think of the analytic beams defined in this section as representing the response of an idealized single feed to an unpolarized sky model.

\subsubsection{CST Beam Interpolation}
\label{subsec:beamint_tests}

\begin{figure}
    \centering
    \includegraphics[width=\columnwidth]{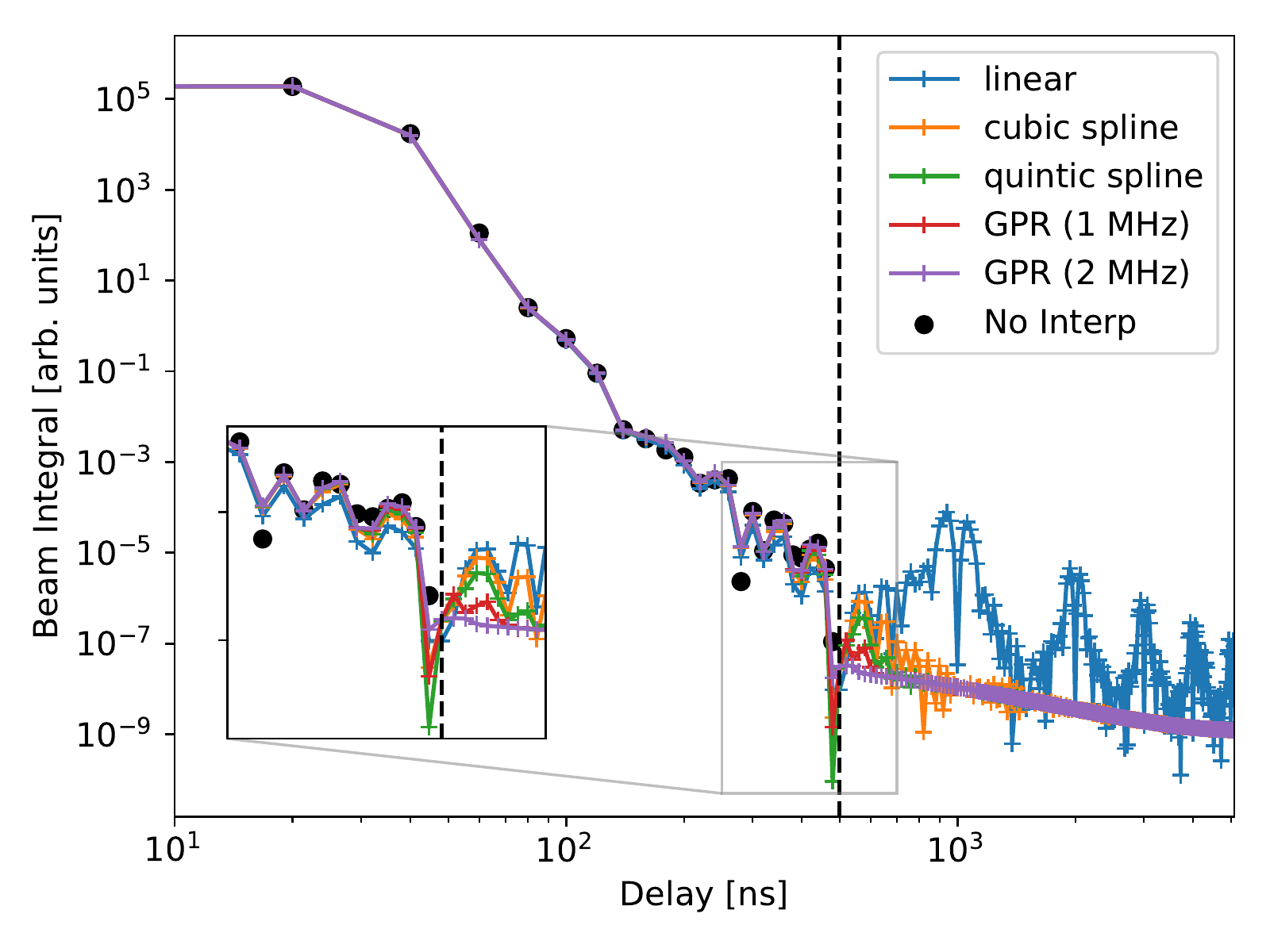}
    \caption{Comparison of the delay-transformed beam integral of the (black dots) non-interpolated beam to the interpolated beam with five methods of interpolation. The vertical dashed line marks 500~ns, the Nyquist sampling rate of the 1~MHz resolution beam. The inset zooms in on the region around the Nyquist limit. There is some disagreement among the curves at and just before this limit, so it seems the effects o interpolation are not strictly limited to interpolated delay modes.}
    \label{fig:beamint_compare}
    % Note --- This figure is made with the script _compare_ints_to_nonint.py in data/NickFagnoniBeams on Oscar.
\end{figure}

Our simulations are done at much higher spectral and angular resolution than the CST simulations, so we need to interpolate the CST simulated beam models values to the required angles and frequencies. The beam data is handled using the new \texttt{pyuvdata} class UVBeam \citep{hazelton_pyuvdata_2017}.\footnote{https://github.com/RadioAstronomySoftwareGroup/pyuvdata} This class supports frequency interpolation at each pixel using the methods accessible to the \textsc{scipy.interpolate.interp1d} method, which includes linear, polynomial splines, and nearest-neighbor interpolation among other methods. UVBeam handles angular interpolation using bivariate spline interpolation.

Interpolating to a higher frequency resolution means adding information in the beam model on higher delay scales. Since the main results of this paper are to measure the behavior of foreground power in delay space, we need to be sure that the frequency interpolation does not have a substantial effect on simulated data. In this section, we interpolate the UVBeam to a finer frequency resolution using a variety of interpolation methods and compare these interpolated beams to the original beam data under a delay transform.

The power beam is obtained from the E-field using \cref{eqn:uvbeam_power}, then the values at each frequency are normalized to their peaks. The peak-normalized power beam is then, at each pixel, interpolated to the 50~MHz bandpass (see \cref{tab:sim_params}) used for simulations. There are two general types of interpolation used: polynomial spline interpolation, which we do to linear, third, and fifth order, and Gaussian process regression (GPR) using the \textsc{scikit-learn} package. The GPR method uses a radial basis function (RBF) kernel, also known as a squared exponential, as the covariance function of the Gaussian process.
For a fixed point on the sky, we can describe the covariance of the beam at two frequencies $\nu_1$ and $\nu_2$ as
\begin{align}
\centering
\label{eq:rbf}
k(\nu_1,\nu_2|\ell) = \exp\left[-\tfrac{1}{2}(\nu_1 - \nu_2)\ell^{-2}(\nu_1 - \nu_2)\right],
\end{align}
where $\ell$ is a hyperparameter that sets the characteristic scale of the covariance function.
Given a covariance function, the GPR maximum a posteriori estimate of the underlying signal constrained by the data can be calculated at any new frequency $\nu^\prime$ \citep[Eq. 2.23 of][]{Rasmussen2006}.
The RBF kernel has the property of producing smoothly varying estimates with frequency structure set by the length-scale hyperparameter.
By keeping this length-scale fixed (rather than solving for its optimal value given the data), the RBF kernel can act as a low-pass filter, filtering out structure at scales above the length-scale \citep[e.g.][]{Kern2019a}. 
Given the inherent Nyquist sampling rate of the CST output, we set $\ell=1$ MHz for our GPR interpolation method.
Altogether, this leaves us with five methods of interpolation to test.

To test the effect of interpolation on the beam, we construct a single metric to compare in delay space. At each frequency, the interpolated beam values are summed over pixels. This list of 512 sums is then is multiplied by a Blackman-Harris window function and then Fourier transformed from frequency to delay using an inverse DFT. The result of these steps is a quantity analogous to the power spectrum measured at $k_\perp = 0$. \Cref{fig:beamint_compare} shows this quantity for each of the interpolation types. The GPR results in the plot are shown with the chosen smoothing scales in parentheses. Overall, all of the interpolation schemes preserve the power on the known scales, marked with black dots, up to the Nyquist limit (shown by the vertical dashed line). Beyond that, the spline interpolations introduce noticeable structure. Further testing has shown that this is likely due to some sharp features in the spectra of pixels far off from the beam center, which can cause aliasing under Fourier transform. The GPR-interpolation seems to smooth out these effects well, and so the curves for the GPR methods are relatively smooth beyond the Nyquist limit.

To avoid any potential issues that may depend on the beam interpolation scheme, we avoid using delay modes beyond the Nyquist limit of 500~ns for any results. Simulations are still carried out to the full frequency resolution of HERA, with beams interpolated using the GPR method with a 1~MHz smoothing scale. This is a conservative choice, because the smoothing scale avoids accidentally smoothing over sub-Nyquist scales in the beam while suppressing the interpolation artifacts. It is of course possible that the HERA beam has structure on scales smaller than 1~MHz, but that cannot be determined from the existing CST simulation data.

\subsection{Sky Models}
\label{subsec:sim-sky_models}

% \begin{figure}
% \includegraphics[width=\linewidth]{gsm_spectral_indices_histo.eps}
% \caption{Histogram of spectral indices fitted to each pixel's spectrum from the GSM model, within the 100 - 120 MHz bandpass used in simulations. Most pixels have spectra near the $-2.5$ expected for synchrotron emission.}
% \label{fig:gsm_indices_histo}
% \end{figure}

\begin{figure}
\subfloat[\label{subfig:gsmplot} GSM]{
\includegraphics[width=\columnwidth]{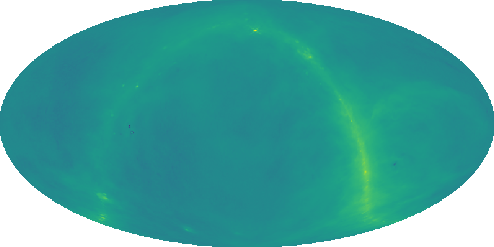}
}\\
\subfloat[\label{subfig:symplot} Symmetric GSM]{
\includegraphics[width=\columnwidth]{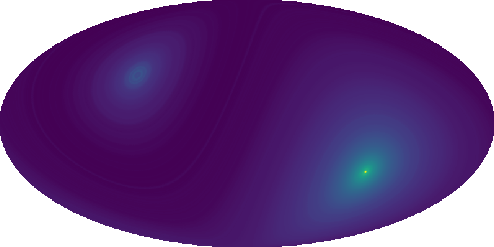}
}\\
\includegraphics[width=\columnwidth]{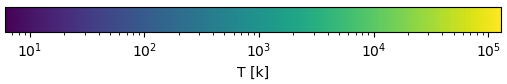}
\caption{The GSM \protect\subref{subfig:gsmplot} and symmetric GSM \protect\subref{subfig:symplot} models in a Mollweide projection of equatorial coordinates.}
\label{fig:mollplots}
\end{figure}

Our simulations model the Stokes-I surface brightness temperature of the sky (in Kelvin) as a set of HEALPix\footnote{Hierarchical Equal-Area Isolatitude Pixellization} maps \citep{gorski_healpix_2005}, one for each frequency channel. HEALPix divides the sky into equal-area pixels with a resolution set by a single parameter, \texttt{Nside}, such that the individual pixel area is $\omega = 4\pi/(12\times \text{Nside}^2)$~sr. With the \textsc{healvis} simulator, there are trade-offs between resolution, run-time, and memory requirements, which are exacerbated by simulating wide fields of view and bandpasses. For this work we use maps at an \texttt{Nside} of 256, which corresponds with a resolution of 13.7 arcmin. The highest resolution probed by any of our simulation configurations is 14.13 arcmin, corresponding with the longest baseline and highest frequency, and shorter baselines are only sensitive to larger scales. \Cref{app:appendix_point_source} shows the effect of increasing resolution on the measured power spectrum of a GSM model is not noticeably improved by increasing the resolution above \texttt{Nside} 256.

We use two diffuse foreground models. The first is the 2016 Global Sky Model (GSM), which is compiled from 29 different sky maps covering a total bandpass of 10MHz to 5TH \citep{zheng_improved_2017}. The GSM is implemented in the Python package \texttt{PyGSM}\footnote{https://github.com/telegraphic/PyGSM}, which generates HEALPix GSM maps at an \texttt{Nside} of 1024 using a principal component decomposition to model the spectral structure at each frequency. We degrade this to an \texttt{Nside} of 256 by averaging together high-resolution pixels nested within low-resolution pixels, which can be done without interpolating or splitting pixel values due to the hierarchical structure of HEALPix maps. In our chosen bandpass, diffuse foreground power is dominated by Galactic synchrotron emission, which generally follows a power law with a spectral index of $-2.5$. Most pixels in the GSM model used here have power law spectra with index $\sim -2.5$.%, as shown in \cref{fig:gsm_indices_histo}.

%We also generate an achromatic version of the GSM by setting the spectrum of each pixel to mean of the GSM spectrum of that pixel. This has the same integrated flux as the GSM, but none of its power-law spectral structure.

The second model is the ``symmetric GSM'' or ``sym'', which is constructed from the GSM but made to have azimuthal symmetry about the Galactic center position. The sym model thus provides a well-defined locus of bright emission on the sky, which will make for a more meaningful test of how spillover relates to the ``position of bright emission''. The sym model is constructed in the following way:
\begin{enumerate}
\item Calculate the angular power spectrum $\mathcal{C}_l$ of the GSM at 100~MHz from the Nside 256 GSM.
\item Generate a set of spherical harmonic expansion coefficients $a_{lm}$, with $a_{l0} = (2l + 1)\sqrt{\mathcal{C}_l}$ and $a_{lm}=0$ for $m \neq 0$.
\item Apply a rotation to the $a_{lm}$ using Wigner D-matrices so that the $(\theta,\phi) = (0,0)$ point is at the Galactic center.
\item Convert the $a_{lm}$ to an Nside 256 map.
\item Scale each pixel in this map by a power law in frequency: $(\nu/\nu_0)^\alpha$ for $\nu_0=$100~MHz and spectral index $\alpha$.
\end{enumerate}
For most tests, we use an achromatic symmetric model, with $\alpha=0$. The angular power spectrum calculation and spherical harmonic transforms are done using \textsc{healpy}'s \texttt{anafast} and \texttt{alm2map} functions, respectively. The rotation is done using \texttt{rotate\_map\_alms}. The GSM and symmetric GSM models at 100~MHz are shown in ICRS equatorial coordinates in \cref{fig:mollplots}, with a log scale on each plot.

\subsection{Power Spectrum Estimation and Measurements}
\label{sec:21cm_pspec_numerical}

The form of the delay spectrum estimator was discussed in \cref{sec:21cm_pspec_general}, in particular with \cref{eqn:pspec_est}. In this section, we will briefly discuss the steps in of estimating power spectra from simulated visibilities, and define the power spillover and fiducial EoR amplitude that will be used to quantify the spread of power beyond the horizon. Rather than look at the power at specific $k_\parallel$ modes, we are interested in the range of $k_\parallel$ that are contaminated by foreground power.

The simulator provides visibilities $V[b,t,\nu]$ for each baseline ($b$), time step ($t$), and frequency ($\nu$), in units of Jy, as well as the beam integral $\Omega_{pp}$ at the zeroth frequency in units of steradians. The visibilities are converted from Jy to mK sr by weighting each channel with a frequency-dependent conversion factor:
\begin{equation}
    V_\text{mK sr}[b,t,\nu] = 10^{-20} \left( \frac{c^2}{2 k_B \nu^2} \right) V_\text{Jy}[b,t,\nu]
\end{equation}

For each baseline and time, the delay transform is done by weighting each channel with the spectral window function and then doing an inverse discrete Fourier transform (IDFT), and taking the absolute square:
\begin{align}
    \widetilde{V}[b, t, \tau_q] &= \frac{B}{N_f}\sum\limits_p \phi(\nu_p) V_\text{mK} e^{2\pi i \nu_p \tau_q}\\
    P[b, t, k_\parallel] &= \frac{X^2 Y}{B_{pp} \Omega_{pp}}|\widetilde{V}[b, t, \pm \tau_q] |^2
\end{align}
In the above, $\nu_p$ is the p$^\text{th}$ frequency, and $\tau_q =  q/(2 B)$ for bandwidth $B$ are the delay modes. Note that there are two delay modes for every $k_\parallel$, since the IDFT goes to both positive and negative delays. Since the power spectrum is a real quantity, the positive and negative delay modes are equivalent, so we average them together.

This leaves us with an estimator of the power spectrum for each baseline, time, and $k_\parallel$. We average together power spectra from baselines with similar lengths (to within a half meter). Additional tests show no significant variation in the power spectrum with baseline orientation. For simulations using the Off-Zenith time configuration, averaging power spectra incoherently over the time steps (55 seconds). The spectrum is not observed to change significantly over such a short period.

For comparison with the foregrounds, we consider a fiducial EoR power spectrum amplitude of $P_\text{21} = 29^2$~mK$^2$~Mpc$^3$, which is approximately the amplitude at $k \sim 1$ Mpc$^{-1}$ at the band-center redshift $z \sim 10.4$. This amplitude is derived from a \textsc{21cmFAST} EoR simulation with default settings \citep{mesinger_21cmfast_2011}. The true EoR power spectrum is stronger at smaller $k$ at this redshift, so this is a conservatively low choice for comparison with the foreground power spectra.

Using the fiducial EoR amplitude, we can define the power spectrum \emph{spillover} $\Delta k_\parallel$ as the distance in $k_\parallel$ that the foreground power extends beyond the horizon. More precisely, the spillover is defined as the largest $k_\parallel$ where the power spectrum crosses below a set threshold of $P_{21}$, such that $P(k_\parallel) < P_{21}$ for all $k_\parallel > \Delta k$, less the horizon $k_h$:
\begin{equation}
    \Delta k_\parallel (P_{21}) \equiv \max\left(k_\parallel | P(k_\parallel) = P_{21}\right) - k_h
    \label{eqn:spillover_definition}
\end{equation}
Defining the spillover relative to the horizon makes it easier to compare the results among different length baselines, which have different horizon limits.

\begin{figure}
\includegraphics[width=\columnwidth]{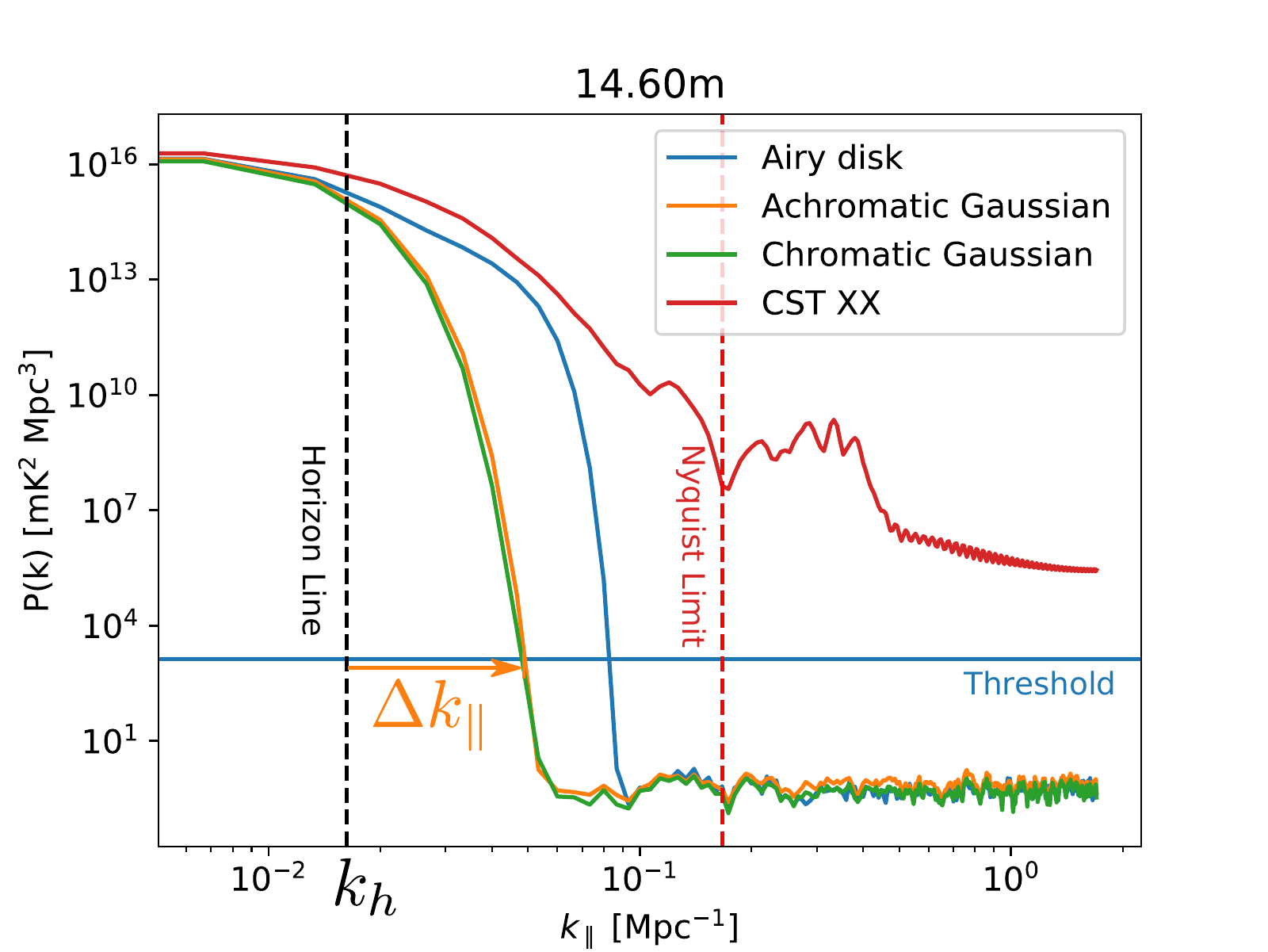}
\caption{The delay spectra of the GSM for a single 14m baseline with a variety of beam types. The fiducial EoR amplitude is shown by the horizontal line, the black vertical dashed line shows the horizon limit, and the foreground power spillover $\Delta k_\parallel$ for the Gaussian beams is indicated by the horizontal green arrow. A vertical dashed red line marks the Nyquist limit for the 1 MHz resolution of the CST beam. For the results with the CST beam, we will use a higher threshold to avoid measuring spillovers past the Nyquist limit. Such measurements would be highly-dependent on the frequency interpolation scheme used.}
\label{fig:bleed_diagram}
\end{figure}

\Cref{fig:bleed_diagram} shows power spectra for a single baseline with several different beam models, with the threshold/horizon/spillover labelled. These spectra are from a simulation with the Off-Zenith time configuration, and with the 14.6~m baselines binned from the hex layout.

\section{Results}
\label{sec:results}
\subsection{Pitchfork Confirmation}
\label{subsec:pitchfork_conf}

The pitchfork foreground signature in Fourier space was first found by simulations in \textsc{PRISim} \citep{thyagarajan_foregrounds_2015}, and later confirmed in data form the Murchison Widefield Array (MWA) experiment \citep{thyagarajan_confirmation_2015}. \textsc{PRISim} was the first interferometer simulator that could handle diffuse sky models down to the horizon, and was thus sensitive to the effects of baseline foreshortening there. \textsc{healvis} is also sensitive to sources near the horizon, and so we expect to see pitchfork power in our simulations. The first set of simulations used a monopole sky signal to confirm the existence of the pitchfork, since a monopole should have minimal power away from the horizon for most baseline lengths.

\begin{figure}
\subfloat[\label{airy}]{\includegraphics[width=\columnwidth]{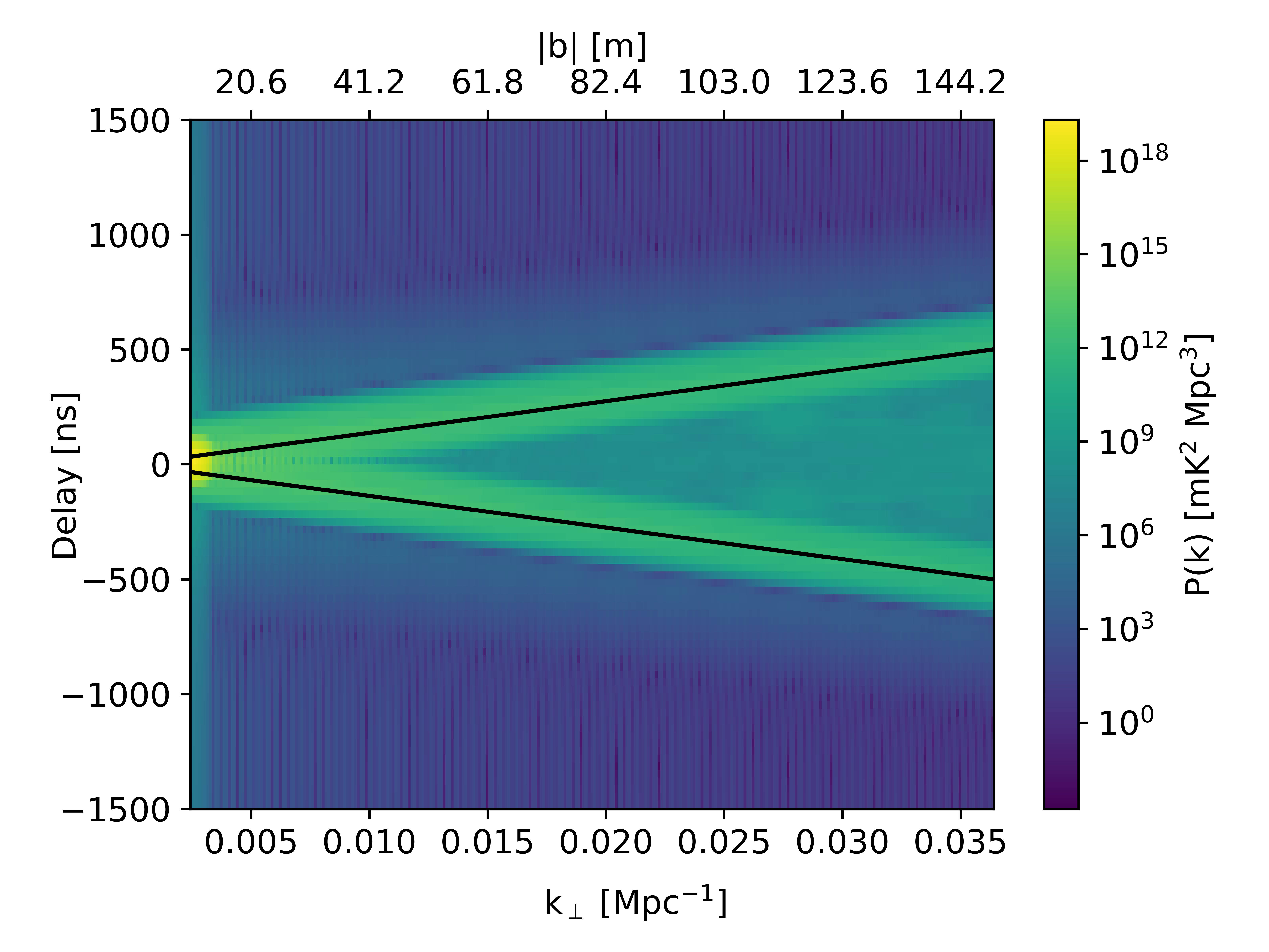}}\\
\subfloat[\label{gauss}]{\includegraphics[width=\columnwidth]{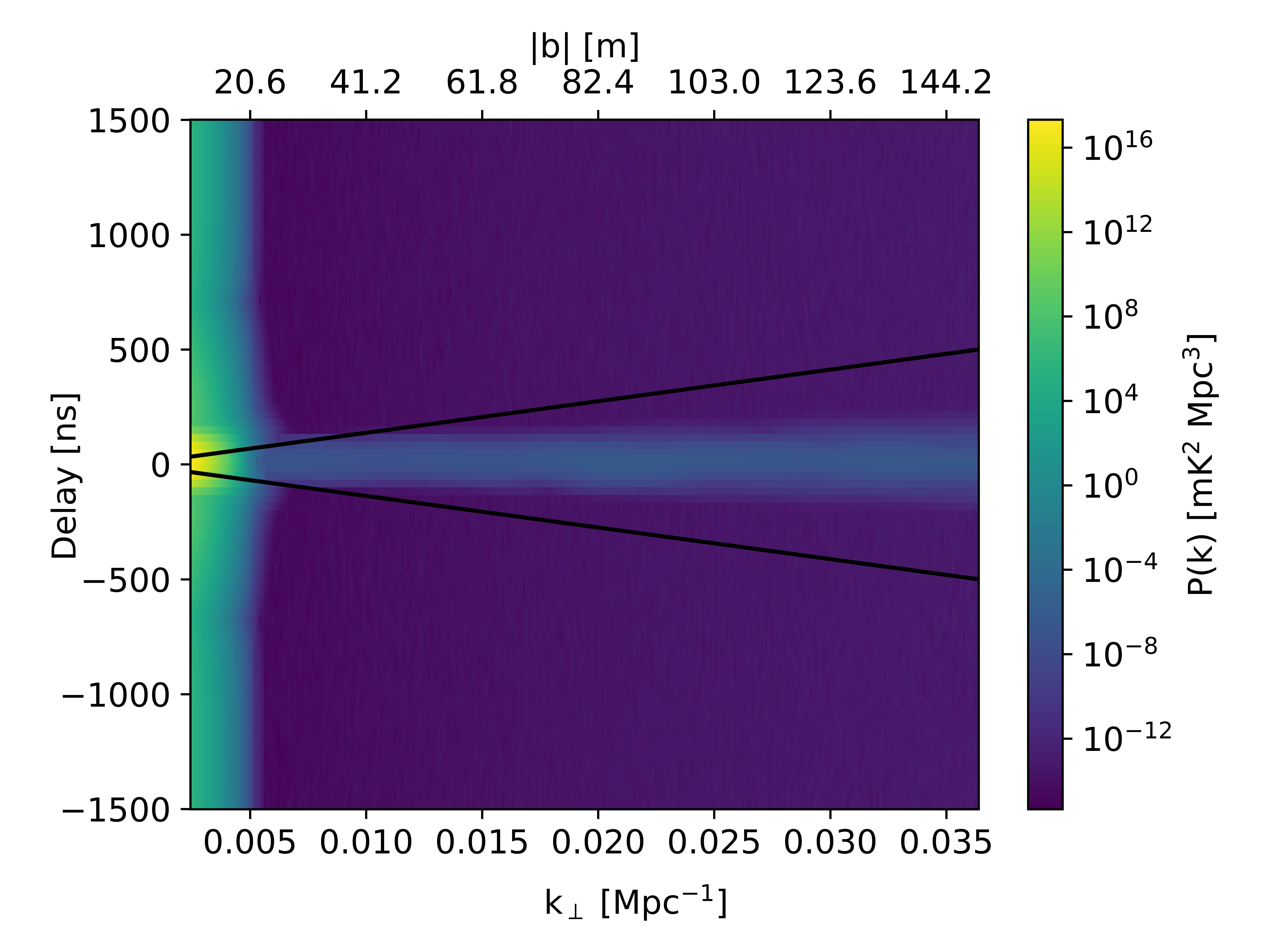}}
\caption{Delay spectra of a monopole signal for a variety of baseline lengths (top axis), and their corresponding $k_\perp$ mode (bottom axis). \protect\subref{airy} shows the results with an Airy beam, which has strong sidelobes and nonzero response down to the horizon. \protect\subref{gauss} is the same for a Gaussian beam, with width set to the width of the main lobe of the Airy beam. Black lines show the delay corresponding with sources on the horizon ($b/c$). Note the difference in color scales.}
\label{fig:pitchfork}
\end{figure}

The simulated array consists of 300 East-West oriented baselines with lengths linearly spaced from 10m to 150m. This provides a range of $k_\perp$ modes to study. We do see the pitchfork for randomly distributed antennas of all orientations as well (see \cref{app:appendix_point_source}). We ran two simulations -- first using the Airy beam, \cref{eqn:airy}, and then with the achromatic Gaussian beam \cref{eqn:agauss}. \Cref{fig:pitchfork} shows the resulting delay spectra. The pitchfork appears for the Airy disk beam, since that has a nonzero response near the horizon. The Gaussian beam is effectively zero at the horizon, and so no pitchfork can be seen. We also observe a strong signal on the shortest baselines. This is evidence that short baselines of an interferometer may be made sensitive to a global signal, as discussed in \cite{presley_measuring_2015}.

The observed pitchfork appears to extend beyond the horizon by a consistent amount for all baseline lengths, suggesting that the pitchfork represents a constant minimum level of foreground spillover for antennas with a nonzero horizon response. Most realistic antennas used in EoR experiments have some response near the horizon, since the width of the beam varies inversely with the effective diameter of the antenna, and most of these experiments use relatively small antenna elements. As we will see, the pitchfork is an important factor of foreground spillover for most of the cases studied here.

\subsection{Window Function Comparison}
\label{subsec:window_func_comp}

As discussed in \cref{subsec:fg_power_spillover}, the shape and width of the bandpass has an effect on the spread of power beyond the horizon, as the intrinsic power spectrum of the source is convolved in delay with the spectral window function and the beam. By choosing an appropriate window function to weight the frequency channels, one can reduce the amplitude of power beyond the horizon at the expense of widening the main lobe of the foreground signal. In this section, we explore the effects of several window functions on power spectra derived from a single symmetric-GSM simulation. The simulations used the Airy beam, Off-Zenith time configuration, and 37 hex layout. The window function is applied to the delay transform for each baseline separately, and the resulting power spectra were averaged over time and binned in $k_\perp$ such that spectra from baselines of the same length to within 10~cm are binned together. The results presented here are for 14.6~m baselines.

The most basic window function is the rectangular window, sometimes called a Dirichlet window, which applies even weighting to all samples in the bandpass. As mentioned, in Fourier space this has the form of a sinc function, which has a limited dynamic range in power. We include it here for comparison, though it is not used in most 21~cm power spectrum analysis.

Note that applying a window function, other than the rectangular, involves down-weighting channels toward the end of the bandpass. This reduces the effective bandwidth $B_{pp}$, and hence increases the width of the main lobe of the Fourier transform, which increases the basic spillover beyond the horizon. There is thus an implicit trade-off between the level of suppression in the foreground sidelobes and the loss of low-k modes to foreground spillover.

We test two window functions, in addition to the unweighted rectangular spectral window. The first is the Blackman-Harris window, which is defined by a sum of three cosine functions optimized to reduce the level of sidelobes in Fourier space (see e.g. \cite{harris_use_1978}). The Blackman-Harris window reduce has a dynamic range of about 120~dB in power, which is generally considered sufficient for 21~cm EoR experiments. \cite{thyagarajan_effects_2016} applied a squared Blackman-Harris window to achieve additional 10 orders of magnitude in sidelobe suppression.

The second window is the Dolph-Chebyshev, which is designed to minimize the sidelobe amplitude for a choice of main lobe width and thus has a tunable dynamic range \citep{SASPWEB2011}. Functionally, it is constructed from Chebyshev polynomials in Fourier space and then inverse-transformed to direct space. Unlike the Blackman-Harris and rectangular windows, the sidelobes do not drop off with $k$, and so usually appear as flat lines in our power spectra. We use Dolph-Chebyshev windows set to 120~dB, 150~dB, and 180~dB ranges, which we refer to as DC120, DC150, and DC180 respectively.

%% Adding here --- More details on the Blackman-Harris and Dolph-Chebyshev windows. Formula relating the width of the DC lobe to the suppression.

\begin{figure}
\includegraphics[width=\columnwidth]{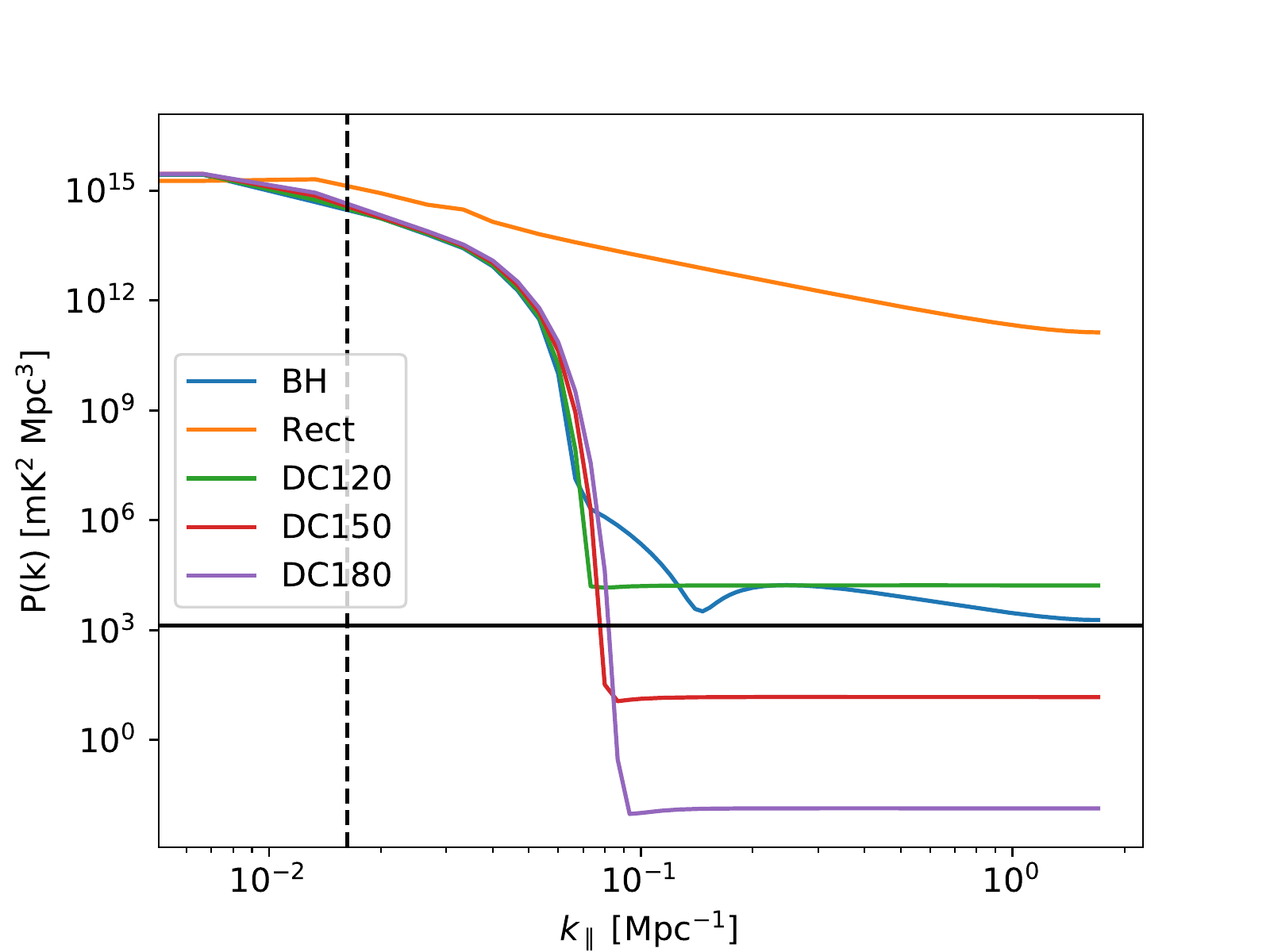}
\caption{Power spectra with rectangular (rect), Blackman-Harris (bh) window functions, as well as Dolph-Chebyshev window functions with dynamic range set to 120~dB and 150~dB (DC120 and DC150, respectively). The horizontal black line marks a fiducial EoR amplitude (see \cref{subsec:window_func_comp}). The vertical dashed line marks the horizon for a 14.6~m baseline. The Blackman-Harris function alone does not provide sufficient dynamic range to push the leaked foreground power below the EoR level. By contrast, applying the right Dolph-Chebyshev window can suppress the leaked foreground power below the EoR threshold.}
\label{fig:window_funcs}
\end{figure}

% Note that these spectra are for a diffuse model only, lacking the contributions of point sources, and so the power spectra of all foregrounds will be higher.

\Cref{fig:window_funcs} shows the comparison of power spectra with the different window functions. The rectangular window function, which has the form of a sinc function in delay space, does not have sufficient dynamic range to keep spilled foreground power below the expected EoR amplitude. The three Dolph-Chebyshev windows demonstrate the expected tradeoff between main lobe width and sidelobe suppression, as can be seen in the slight spread among the curves around $P(k) \sim 10^{9}$ mK$^2$ Mpc$^{3}$). The Blackman-Harris window displays similar behavior to the DC120 window, and continues to drop off at higher $k_\parallel$ while the Dolph-Chebyshev window function stays flat.

For the rest of this paper, we use the DC180 window for all delay spectrum estimation. The Dolph-Chebyshev is the only window tested that has sufficient dynamic range to reach the EoR amplitude without any assumed foreground subtraction,\footnote{the Blackman-Harris window does drop below this flat EoR threshold at the high $k_\parallel$ for some limited cases, but not enough to be useful for our results} and does so with comparable foreground spillover to the Blackman-Harris used in other work (see, e.g., \cite{thyagarajan_study_2013, parsons_sensitivity_2012, vedantham_imaging_2012}).

\subsection{Spectral Index Comparison}
\label{subsec:spec_index_comp}

%[NOTE -- The total flux is changing for sky models with spectral indices. I may need to re-scale them so that they share the same DC mode for the comparison to be useful.]

\begin{figure}
\includegraphics[width=\columnwidth]{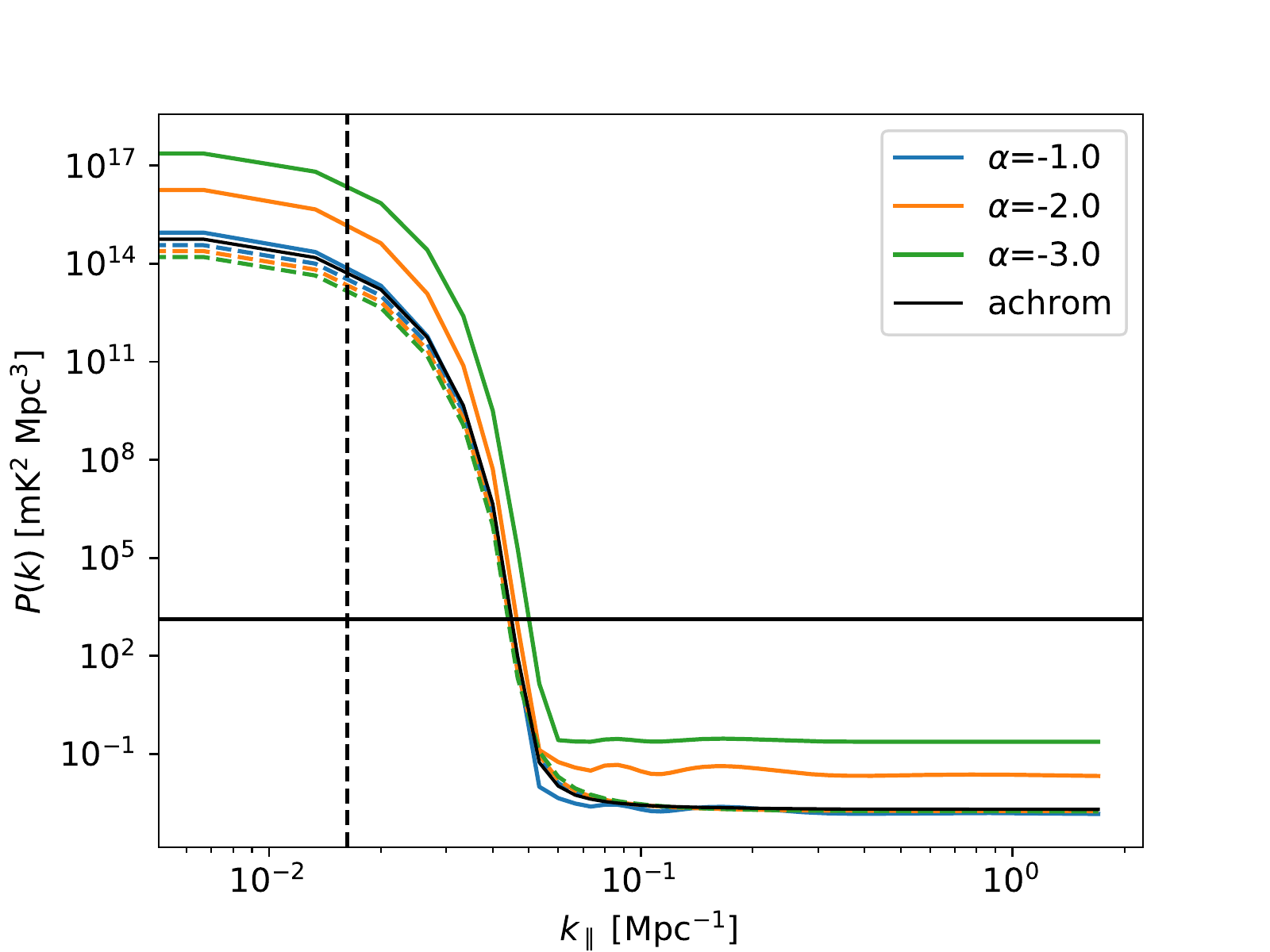}
\caption{Power spectra of a single 14.6~m baseline with a Gaussian beam. The color indicates the spectral index of either the beam or the sym sky model. For solid lines, the spectral index is applied to the beam and the sky model is flat in frequency. For dashed lines, the spectral index is applied to the sky model and the beam is flat-spectrum. The black line shows the result of an achromatic beam and sky model.}
\label{fig:pspec_v_specgauss}
\end{figure}

The next tests explored the effects of power law spectral structure in the primary beam to structure in the sky model itself. As mentioned in \cref{subsec:sim-sky_models}, diffuse power typically has a power law spectrum. The angular width of a beam typically evolves linearly with frequency, as for the Airy disk. Here we look to slightly steeper evolution of the beam width and sky brightness. Note that the structure is applied to the model an the beam in slightly different ways -- the the beam is normalized to its peak value at all frequencies, but its width changes with frequency. The sky model's total brightness is evolving with frequency instead.

We ran simulations with the following configurations:
\begin{itemize}
\item Chromatic Gaussian beam with power law index $\alpha=\lbrace -1, -2, -3\rbrace$, Flat-spectrum sym model. (Solid lines in \cref{fig:pspec_v_specgauss}).
\item Achromatic Gaussian beam, sym model with spectral index $\alpha=\lbrace -1, -2, -3 \rbrace$. (Dashed lines in \cref{fig:pspec_v_specgauss})
\end{itemize}

These simulations were run with the 37 hex antenna layout with the Off-Zenith times (see \cref{tab:sim_params}). Power spectra were averaged in time and binned in baseline length. \Cref{fig:pspec_v_specgauss} shows the power spectra of 14.6~m baselines for these different sky and beam models. Solid lines are for simulations with a chromatic beam, while dashed lines are for simulations with the spectral sky model. The amplitude at low-k shifts with spectral index for a couple of reasons. When the spectral index is applied to the sky model, the model has less power at higher frequencies as compared with the flat-spectrum model, so the total power drops. When the spectral index is applied to the beam width, the beam is narrower at higher frequencies. A narrower beam in image space corresponds with a wider footprint in Fourier space, and hence integrates more power. This causes the low-k modes of the solid curves to shift higher with decreasing (increasingly negative) spectral index.

The effects of the beam chromaticity are much more noticeable, shifting the overall amplitude higher, than the skymodel chromaticity. Since the diffuse foreground power is known to largely follow a clean power law within this range, we take this to mean that the spectrum of the sky has a negligible effect on the foreground spillover. Increasing beam chromaticity does also shift the spillover out to slightly higher $k_\parallel$, although not to the degree that can be seen in a more complicated beam response as with the CST beam as seen in \cref{fig:bleed_diagram}.

%From \cref{fig:pspec_v_specgauss}, it seems that the sky model chromaticity has a stronger impact on the power spectrum than the spectral structure of the beam, but neither has an especially dramatic effect on the distance the power spectrum shifts beyond the horizon line.

\subsection{Foreground Spillover}
\label{subsec:bleed_results}

The foreground spillover is estimated by fitting a polynomial spline to the power spectrum measurements and finding the points where the spline crosses the threshold. The spline fit is done in log-log space where it is most stable. The largest $k_\parallel$ of intersection, less the $k_\parallel$ of the horizon, is taken as the measured foreground spillover $\Delta k_\parallel$, as defined in \cref{eqn:spillover_definition}.

\subsubsection{Time}
\label{subsubsec:bvt_results}

We ran a set of simulations that covered 24 hours of LST in steps of 5 minutes (the ``transit'' time set), in order to measure how foreground power spillover on single baselines changes with the hour angle of the galactic center (recall that the ``galactic center'' also refers to the center of the sym model). As the galactic center moves away from the zenith, the brightest part of the sky is shifted toward a weaker part of the primary beam, but the delay spectrum peak moves closer to the horizon. Brightness near the horizon is further enhanced on all baselines by the pitchfork effect, so it is interesting to see how the power spillover vs. time changes with different baseline lengths.

%We fit a polynomial spline to the power spectrum curve in log-log space, and then estimate the maximum $k_\parallel$ at which the spline fit intersects crosses the threshold. This gives a minimum $k_\parallel$ such that the power spectrum is less than the threshold for all larger $k_\parallel$. Our $\Delta k$ estimates are therefore conservative upper limits for the spillover, beyond which foreground contamination is lower than the chosen threshold.
\begin{figure}
\subfloat{\includegraphics[width=\columnwidth]{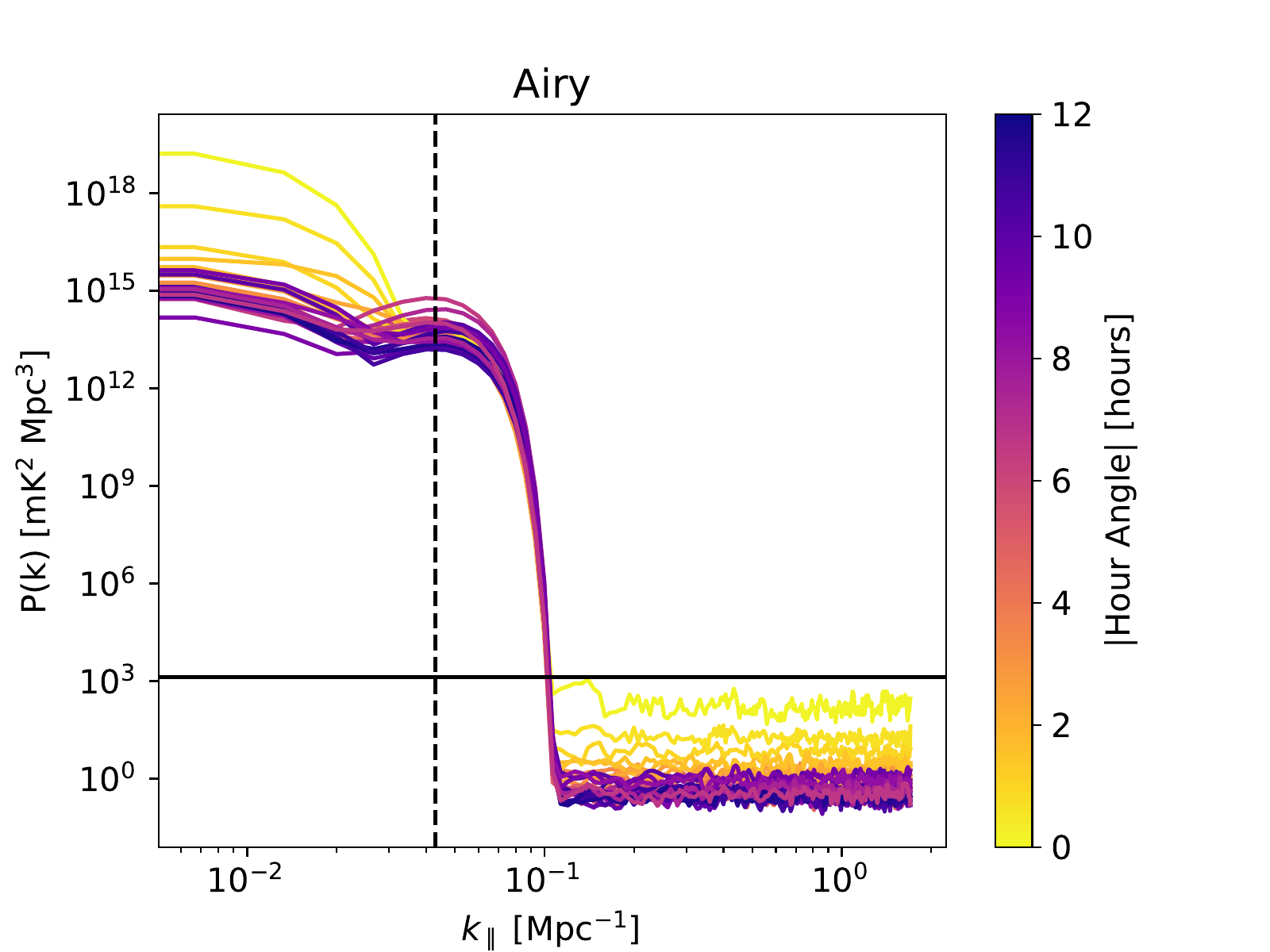}}\\
\subfloat{\includegraphics[width=\columnwidth]{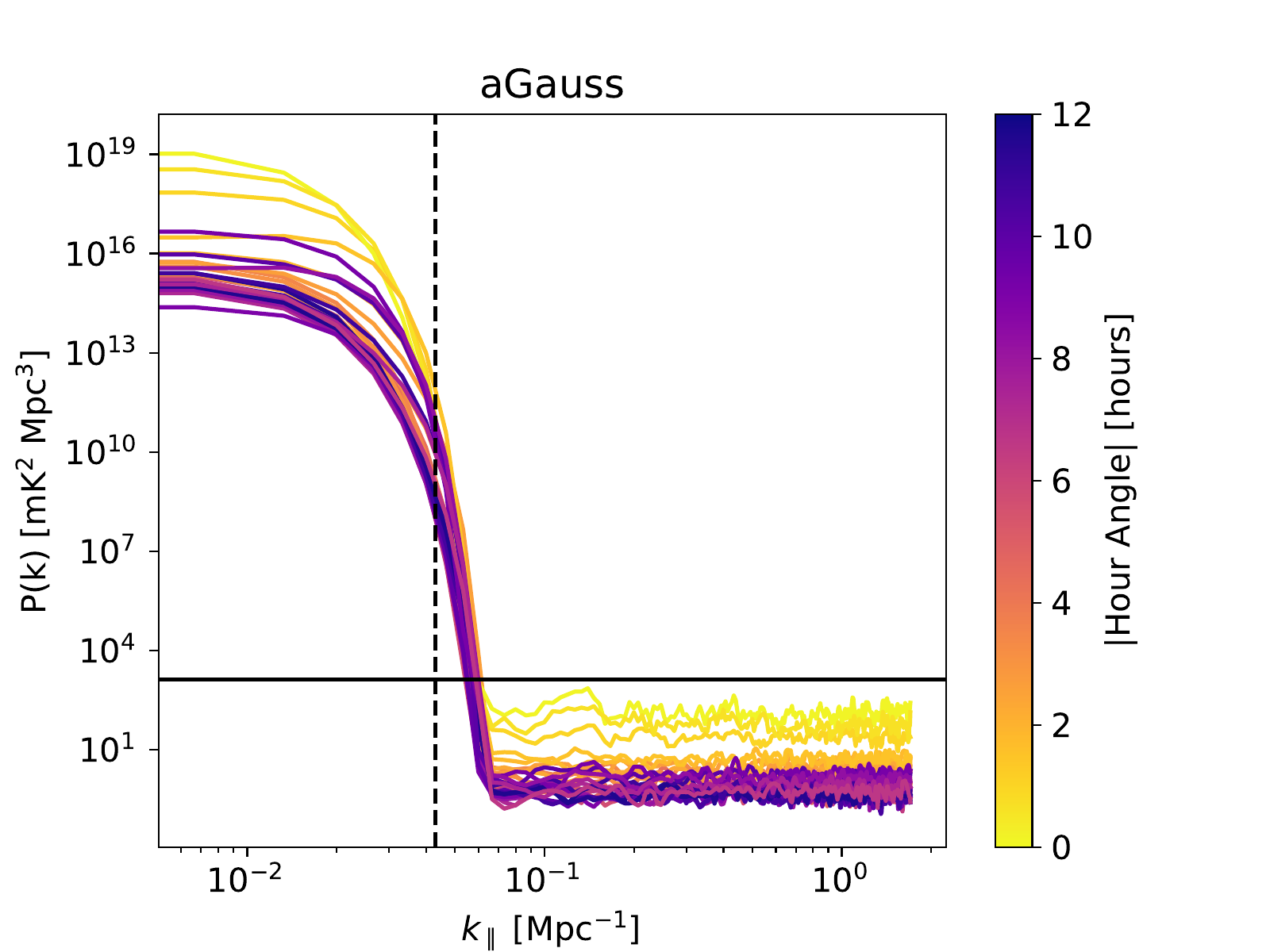}}
\caption{1D power spectra for a 38.6~m baseline vs. the absolute value of the galactic center hour angle. Each curve corresponds with a different hour angle of the galactic center, for the symmetric GSM model. The vertical dashed line marks the horizon limit, and the horizontal line marks the fiducial EoR amplitude. The top is for the Airy beam; the bottom is for the achromatic Gaussian beam. There Airy beam shows a persistent ``shoulder'' near the horizon that is due to the foreground pitchfork, while power spectra with the gaussian beam do not. The pitchfork keeps the spillover at the EoR threshold line very nearly constant for all hour angles, despite the changing amplitude within the horizon.}
\label{fig:pspecs_vs_time}
\end{figure}

\Cref{fig:pspecs_vs_time} shows the power spectra vs. galactic center hour angle for a 14.6~m baseline for the symmetric GSM model with Airy (top) and achromatic Gaussian (bottom) beams. The vertical dashed line marks the horizon and the horizontal line marks the chosen threshold $P_\text{thresh} = 29^2$~mK$^2$~Mpc$^3$. There is a clear increase in total power as the galactic center (hence, brightest part of the symmetric sky model) transits overhead. For the aGauss beam, this causes the first sidelobe of the Blackman-Harris window to rise up through the threshold, quickly pushing the $\Delta k_\parallel$ higher. For the Airy beam, the pitchfork effect is apparent as a persistent bump near the horizon line, and the spillover stays relatively constant except for when the galactic center is overhead.

\begin{figure*}
\centering
\includegraphics[width=\columnwidth]{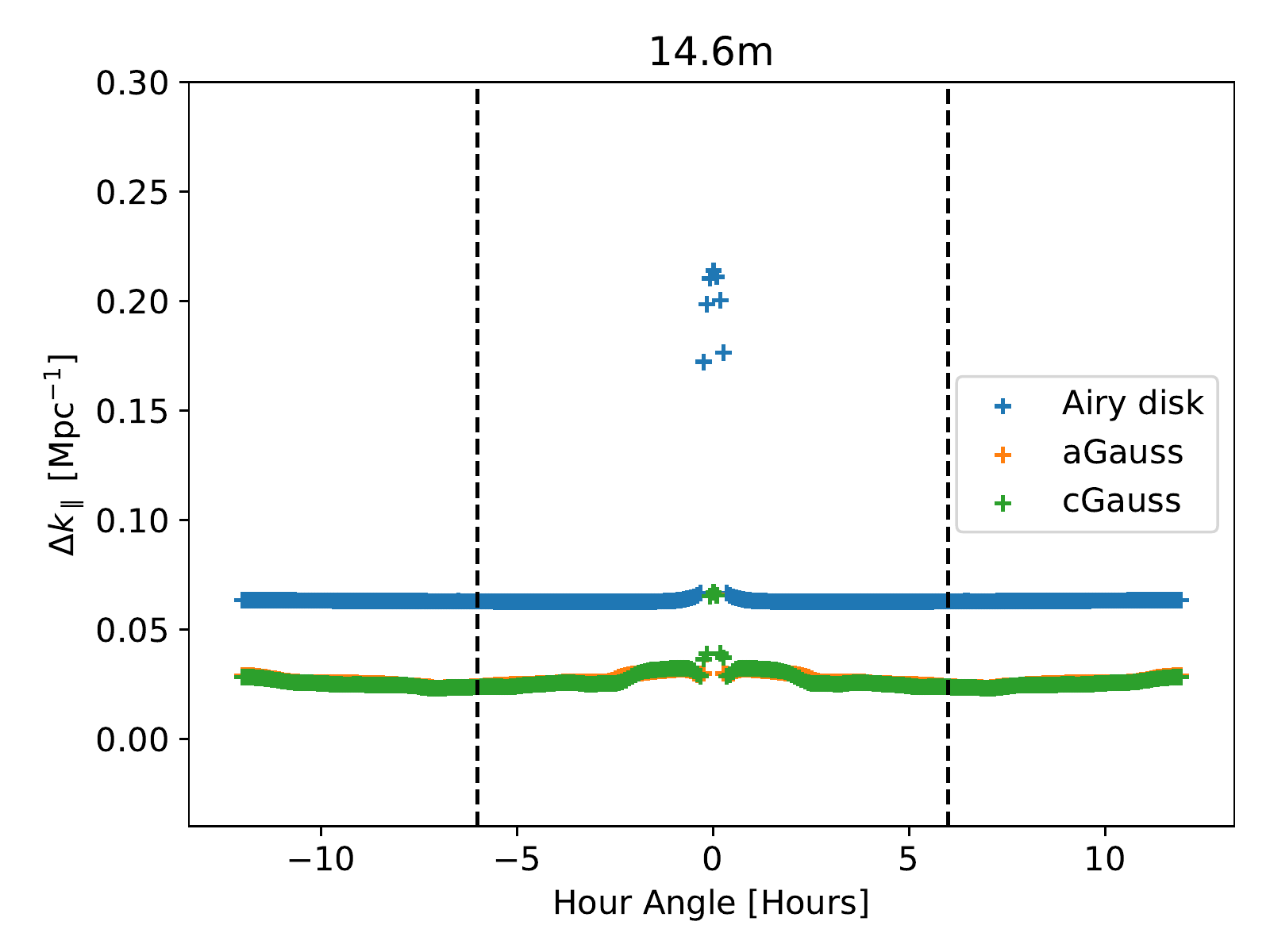}
\includegraphics[width=\columnwidth]{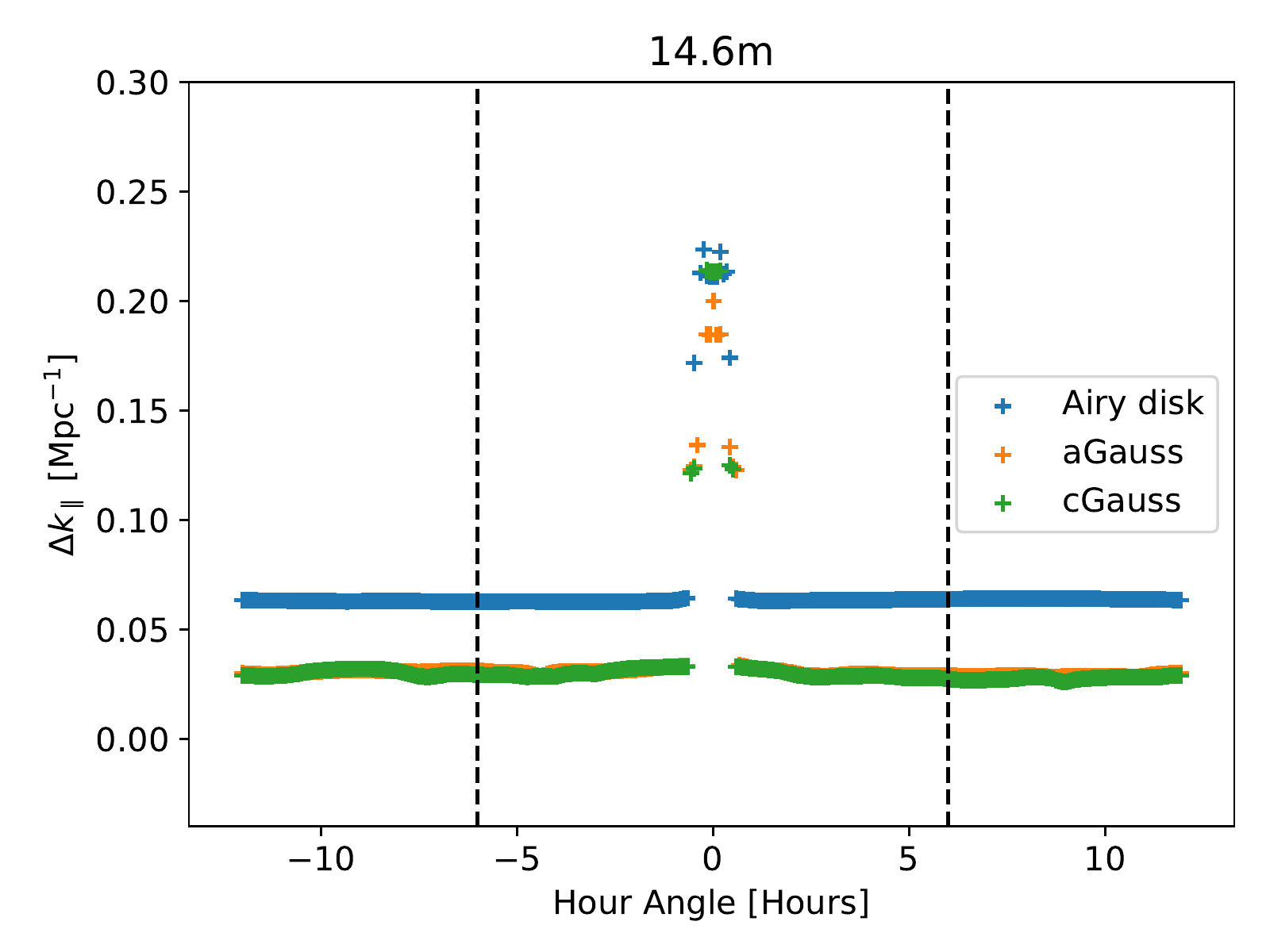}\\
\subfloat[\label{subfig:symgsm_lvt} sym]{
\includegraphics[width=\columnwidth]{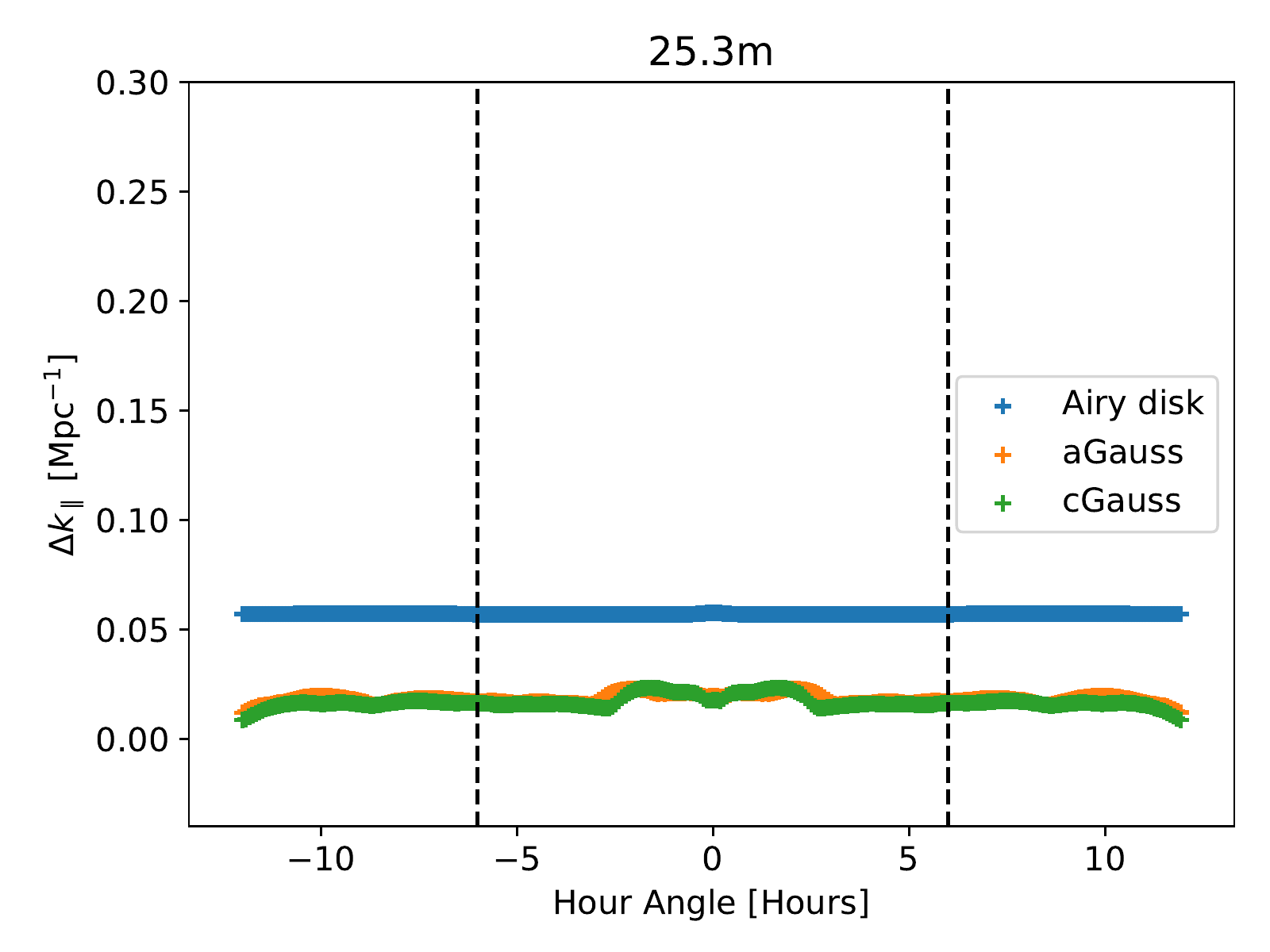}}
\subfloat[\label{subfig:gsm_lvt} GSM]{
\includegraphics[width=\columnwidth]{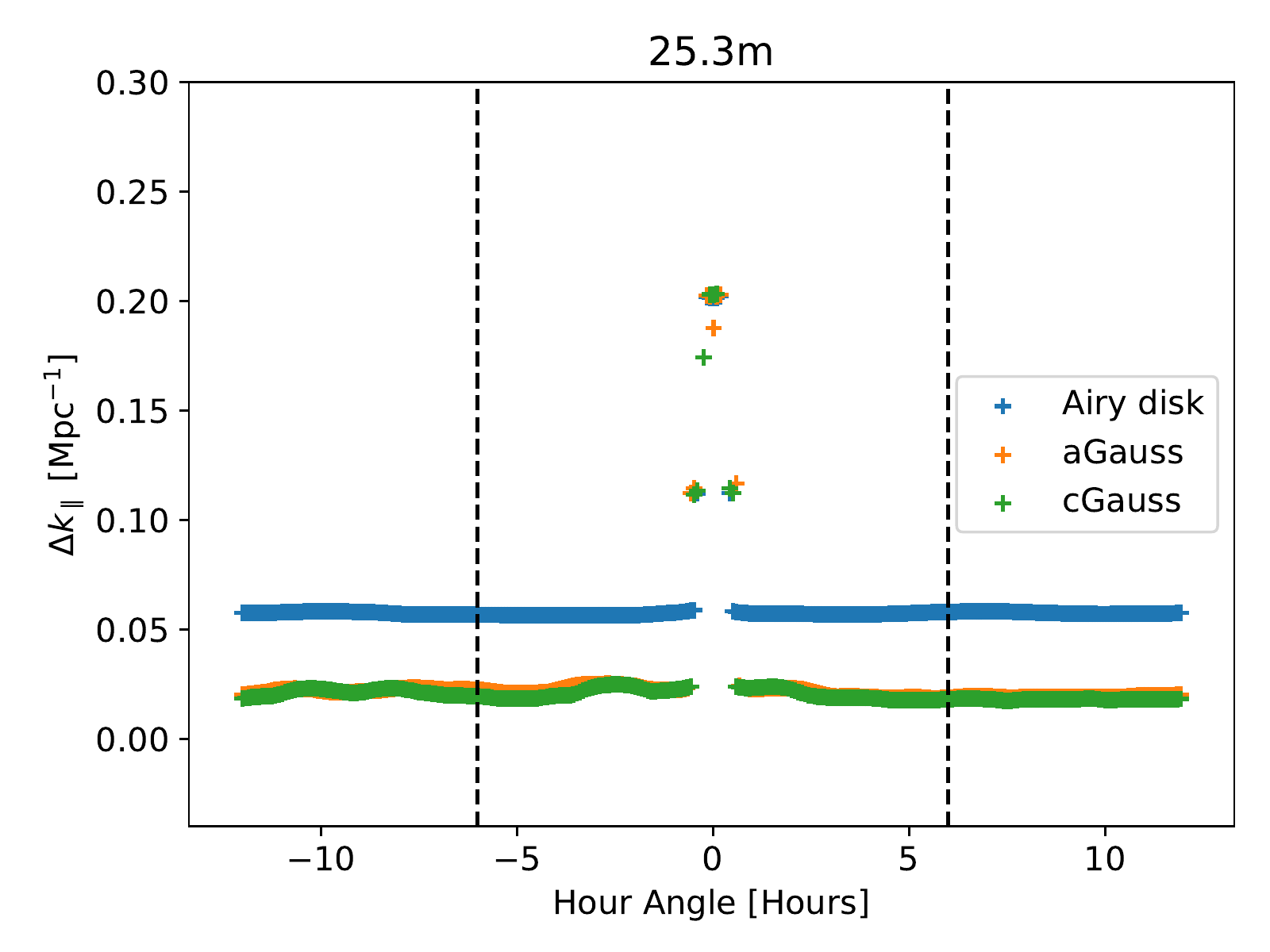}
}
\caption{The foreground spillover at the threshold shown in \cref{fig:pspecs_vs_time}, for a 14.6~m baseline (top row), plotted against the hour angle, for the three analytic beams. Vertical dashed lines mark $\pm$6 hours, when the galactic center (or, symmtric GSM center) is rising or setting. The spikes near $\simeq 0$~hrs arises from the Dolph-Chebyshev sidelobe ripples rising above the threshold. Overall, the spillover remains fairly constant with time, with the Airy disk beam showing consistently higher spillover due to the constant presence of the pitchfork. On a longer baseline (bottom row), the overall amplitude is lower since longer baselines collect less power from diffuse emission.}
\label{fig:bleed_v_time}
\end{figure*}

\Cref{fig:bleed_v_time} shows the measured foreground spillover vs. galactic hour angle for the symmetric GSM (\ref{subfig:symgsm_lvt}) and the actual GSM (\ref{subfig:gsm_lvt}) sky models for a 14.6~m baseline. As will be shown in the next section, the spillover is highest on short baselines because the foreground power is strongest there. Vertical dashed lines mark the rise and set times for the galactic center, which in both cases is the brightest part of the sky. The symmetric GSM model shows roughly the same behavior as the GSM, with a few minor differences. The symmetric GSM is flatter and more symmetric with hour angle, as expected. The GSM is less peaked at hour angles near zero, probably because the full power is more spread out across the sky than in the sym model. For the Airy beam, which has a non-negligible response near the horizon and thus has a consistent pitchfork, the spillover is higher. It is clear that the pitchfork dominates suprahorizon emission during times when the galaxy is out of the primary beam. The two Gaussian beams show very similar behavior, with minimal spillover at times when the galactic center is away from the zenith.

\subsubsection{Baseline Length}

\begin{figure*}
\centering
\subfloat[\label{subfig:symgsm_lvb} sym]{\includegraphics[width=\columnwidth]{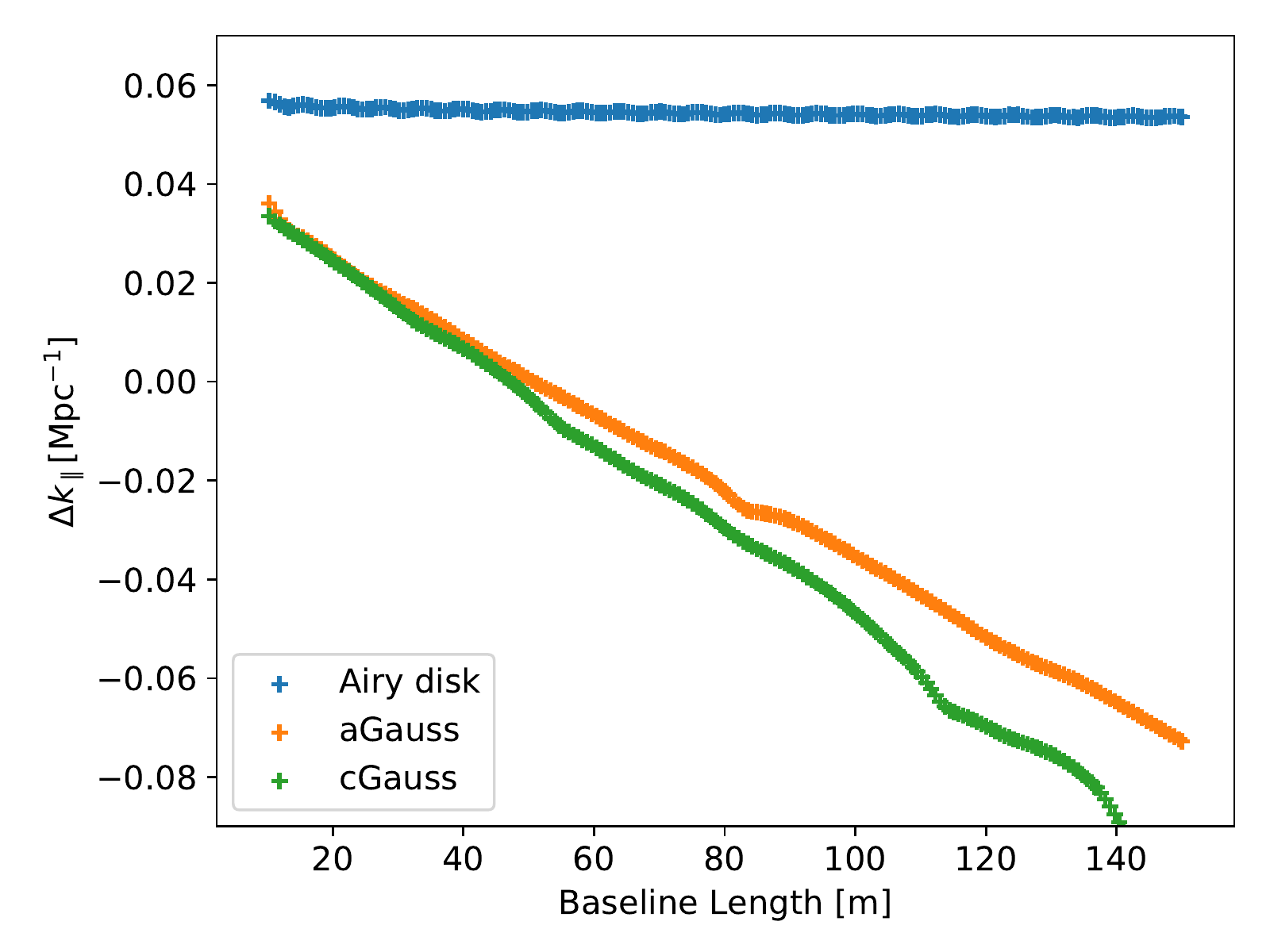}}
\subfloat[\label{subfig:gsm_lvb} GSM]{\includegraphics[width=\columnwidth]{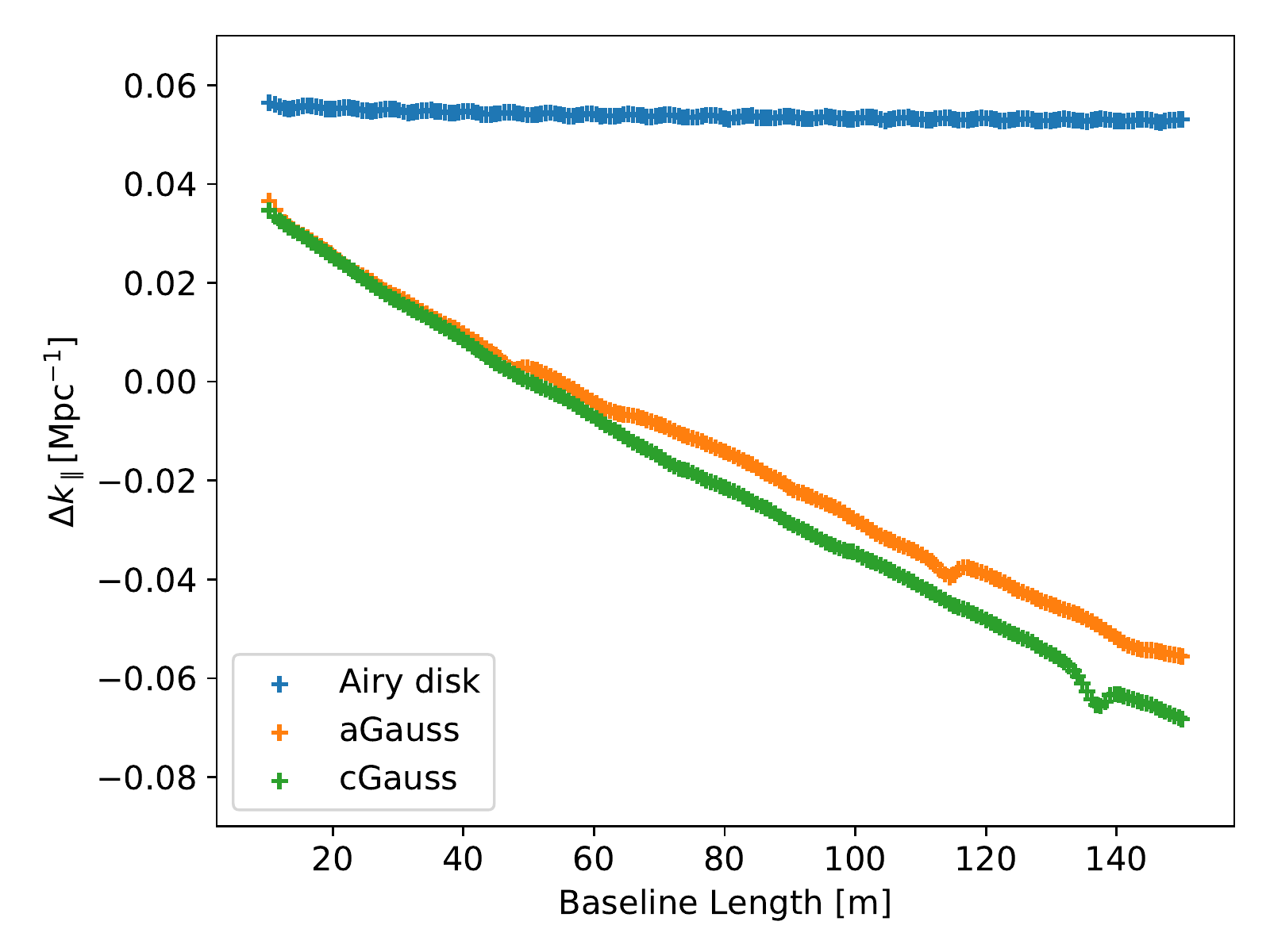}}
\caption{The foreground spillover at the threshold shown in \cref{fig:pspecs_vs_time}, plotted against East-West baseline length, for the three analytic beams. The black line marks the horizon. These simulations were done with the Off-Zenith time configuration, when the spillover is maximized. Spillover increases steadily with decreasing baseline length. Note that a negative spillover means that foreground power is contained within the horizon limit at the EoR amplitude.}
\label{fig:bleed_v_bllen}
\end{figure*}

The last set of simulations looked at effect of baseline length on foreground power spillover. For this set we use the Line antenna layout, which provides 300 East-West baselines with lengths spanning 10 to 150~m (the ``line'' configuration). The 150~m baseline here is the longest baseline used here for foreground spillover measurements. We note that this is still within the range of baseline lengths where the point source approximation is good (see \cref{app:appendix_point_source}). These simulations used the Off-Zenith time set, and power spectra were averaged over that time span.

\Cref{fig:bleed_v_bllen} shows the results with the symmetric GSM (\ref{subfig:symgsm_lvb}) and full GSM (\ref{subfig:gsm_lvb}). In both cases, the foreground spillover is dominated by the pitchfork for the Airy beam above about 50~m, and is pushed up by the first sidelobe of the BH window on shorter baselines. For the Gaussian beams, which have no pitchfork, the spillover simply drops off with increasing baseline length. In all cases the spillover decreases with increasing baseline length, albeit more slowly for the Airy beam, due to the decrease in power measured with longer baselines, as is expected for diffuse models.

\subsection{CST Beam Results}
\label{subsec:cst_beam_results}

\begin{figure}
    \centering
    \subfloat[\label{subfig:gsmcst}GSM]{\includegraphics[width=\columnwidth]{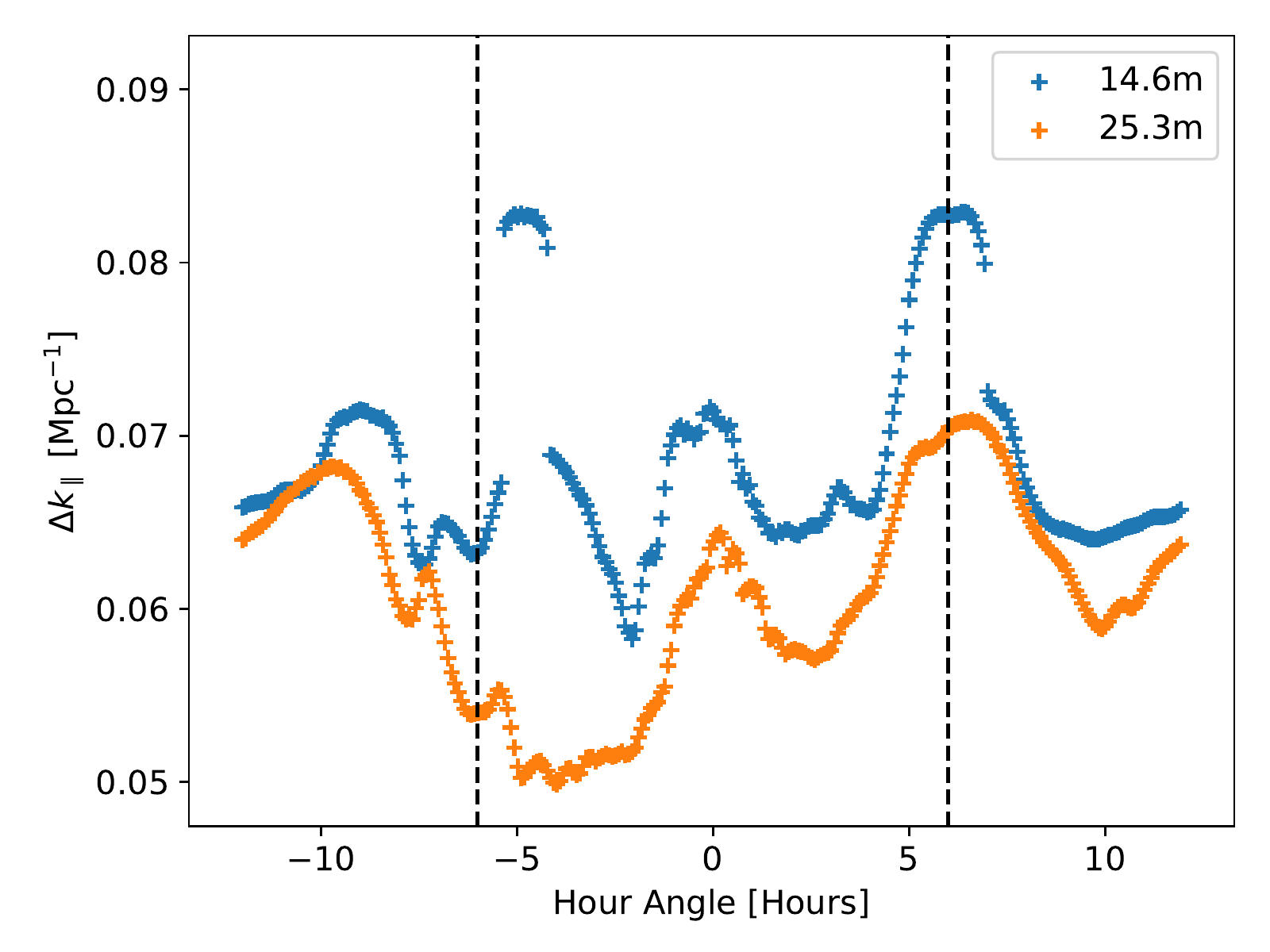}}\\
    \subfloat[\label{subfig:symcst}Sym]{\includegraphics[width=\columnwidth]{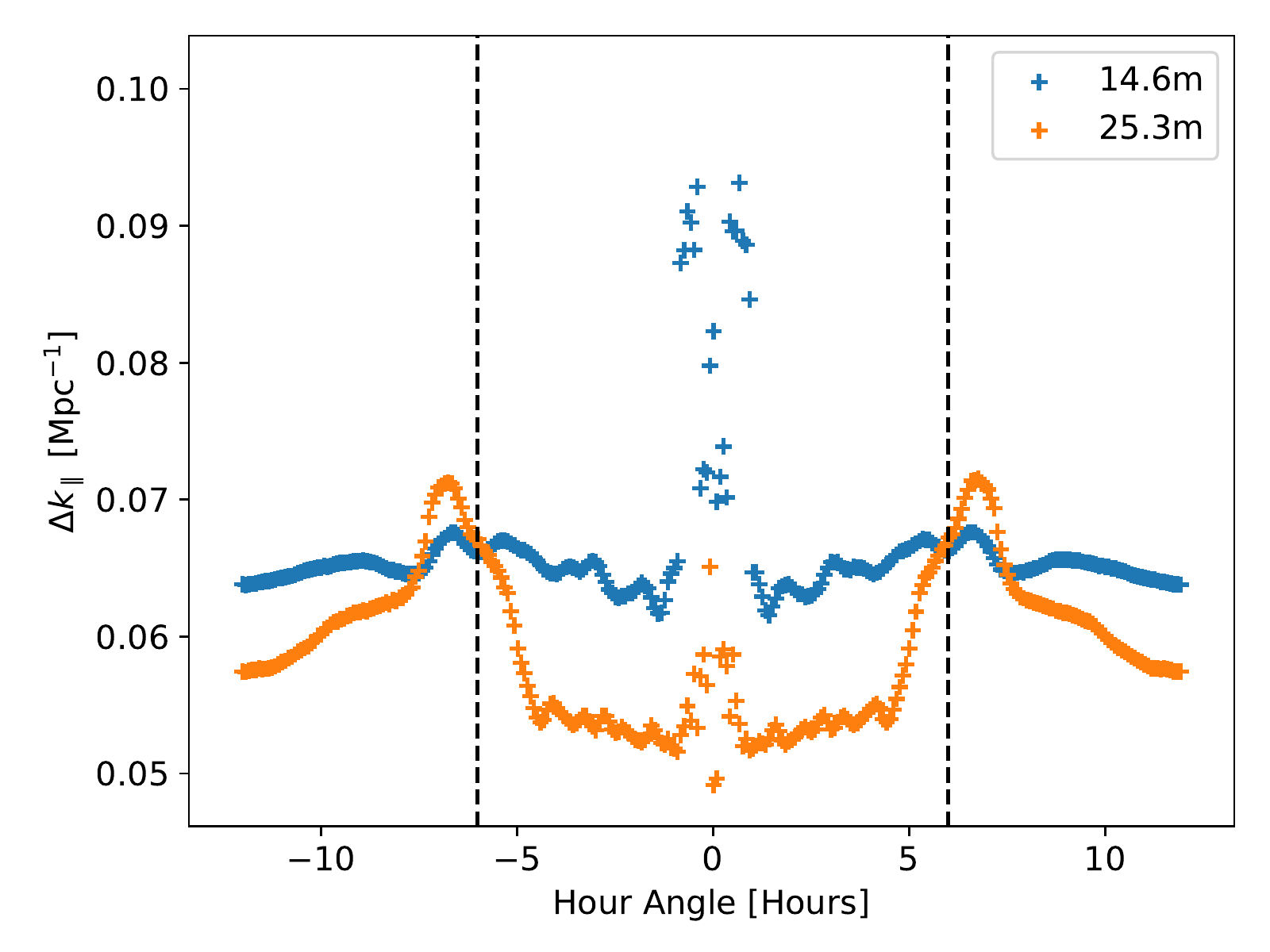}}
    \caption{The spillover vs. hour angle for the \protect\subref{subfig:gsmcst} GSM and \protect\subref{subfig:symcst} sym models, for 14.6~m (blue) and 25.3~m (orange) baselines. Horizon lines are marked in the same colors. The strong response of the CST beam near the horizon causes the spillover to increase when the galactic center is near the horizon, at $\pm$ 6 hours (marked with dashed vertical lines). The longer baseline has less power spillover than the shorter for most hour angles, likely because the total power it measures is less, but at the horizon the curves approach each other. This pattern is especially clear for the sym model.}
    \label{fig:cst_bleed-v_time}
\end{figure}

%The CST beam, as discussed in \cref{subsec:beamint_tests}, introduces non-negligible structure when interpolated in frequency. This structure dips slightly below the Nyquist rate of of the reduced-frequency beam model, as shown in \cref{fig:beam_interp_errs}. For the full-frequency beam, which is sampled at every 1~MHz, we therefore expect to see interpolation artifacts at delays above $\tau \sim 1/(2 \times 10^6)$~sec, or $k_\parallel \sim 0.167$Mpc$^{-1}$ at $z \sim 10.37$. These artifacts are visible in the power spectra plotted in \cref{fig:bleed_diagram}, and their structure extends below this limit a little. This is also the point where the power spectrum dips below the fiducial EoR amplitude.

\begin{figure*}
\subfloat{\includegraphics[width=\columnwidth]{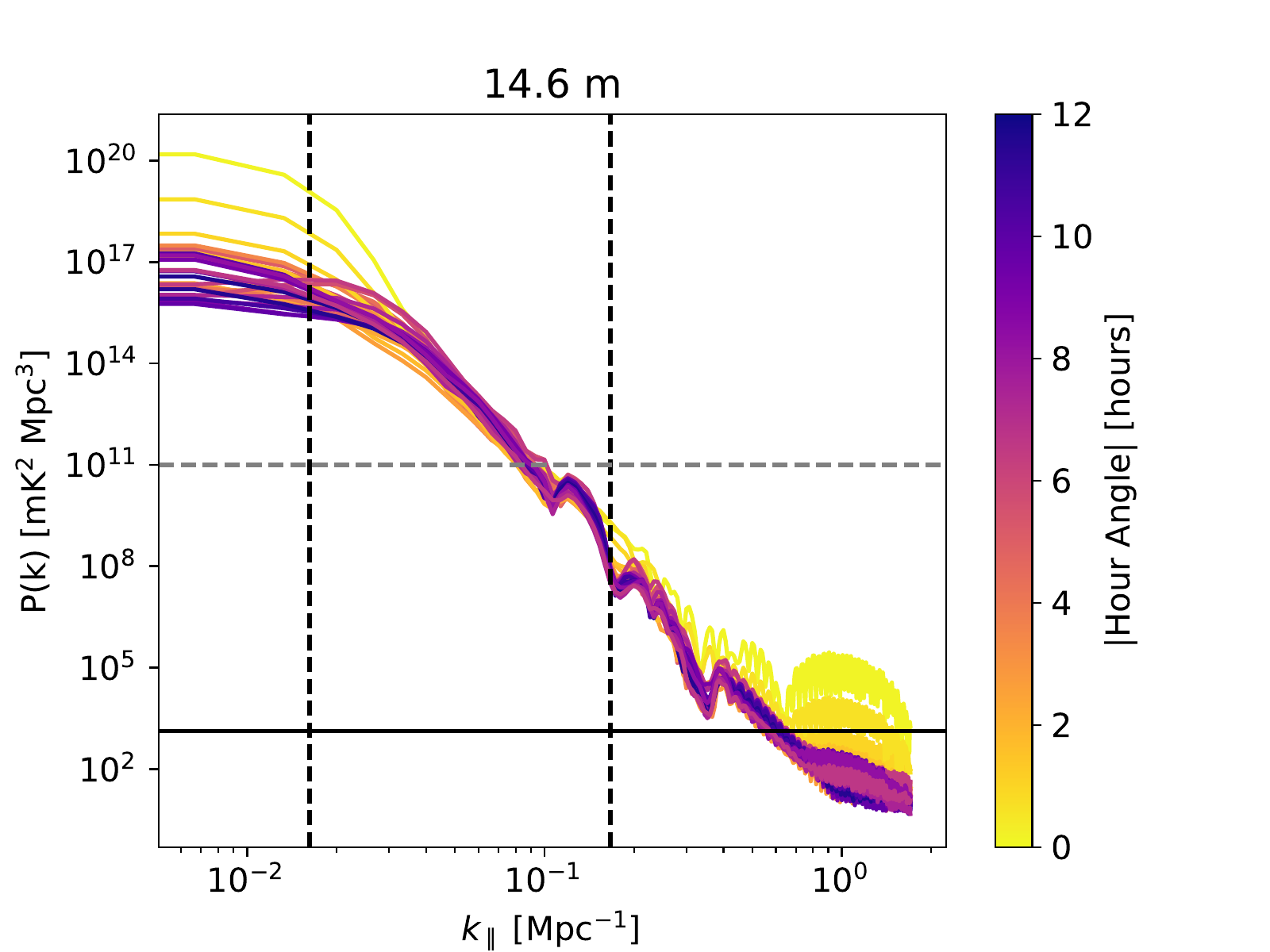}}
\subfloat{\includegraphics[width=\columnwidth]{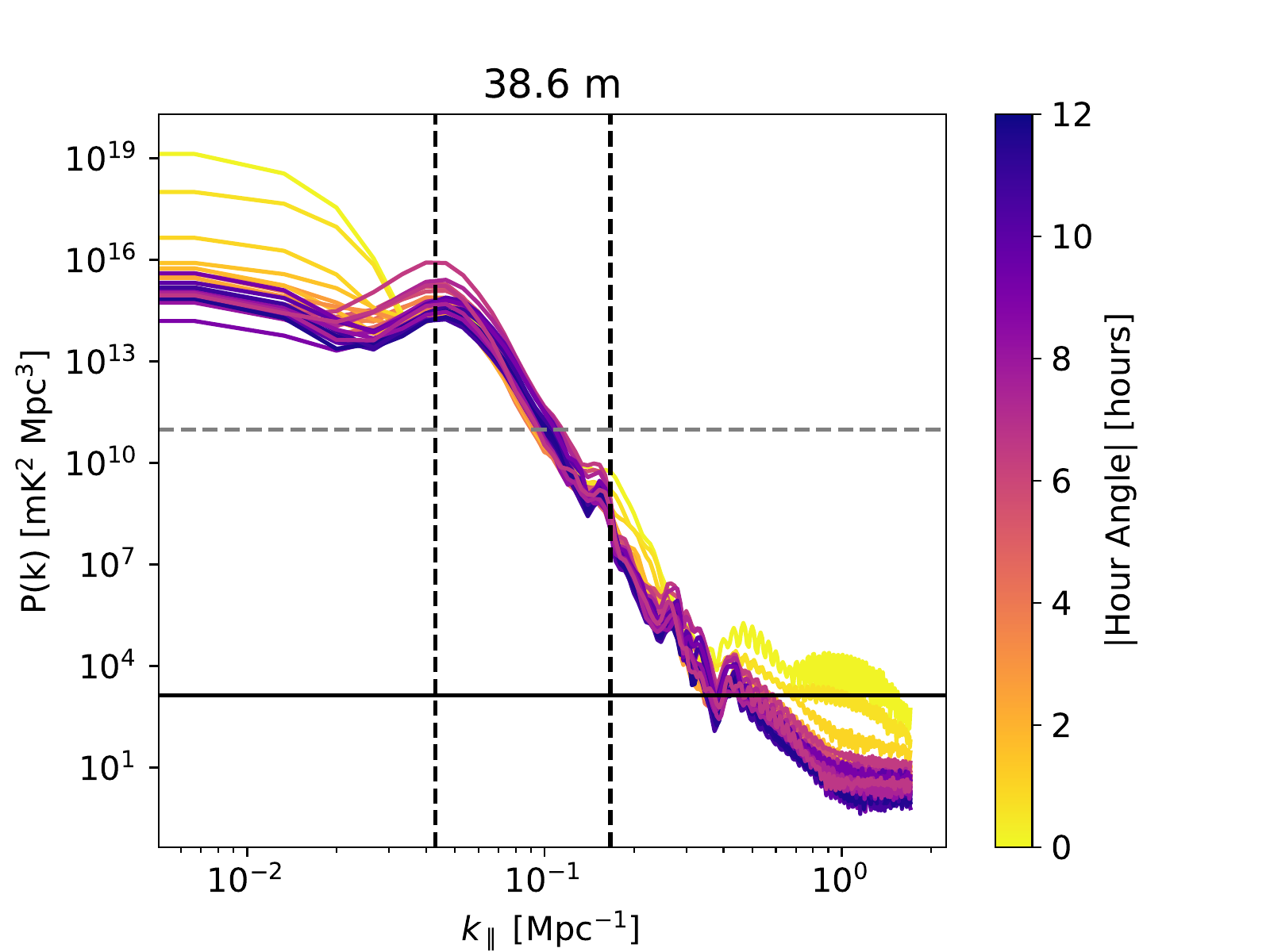}}
\caption{Power spectra vs. galactic center hour angle for the achromatic sym model and the CST XX beam. The vertical dashed lines in each plot are the horizon and Nyquist limit, while the horizontal lines show the EoR amplitude threshold used before (solid, black) and the new threshold for \cref{subsec:cst_beam_results}.}
\label{fig:pspecs_vs_time_cstxx}
\end{figure*}

In this section we share the results of simulations with the CST beam. \Cref{fig:pspecs_vs_time_cstxx} shows the power spectra vs galactic center hour angle for the achromatic sym model with the CST XX beam at two baseline lengths. The pitchfork is present for both baseline lengths, but is clear persistent on the longer baseline at a higher amplitude than the power at smaller $k_\parallel$. As we have seen with the analytic beams, the pitchfork is largely independent of baseline length. Note also that for almost all time, the power spectrum drops below the EoR amplitude threshold (solid black horizontal line) at $k_\parallel$ well past the Nyquist limit. We therefore choose a higher threshold of $10^{11}$~mK$^2$~Mpc$^3$ (the horizontal, dashed, gray line) for measuring foreground spillover with the CST beam. This avoids measuring modes that were only reached by interpolation. Though these results are not at the EoR amplitude, they demonstrate the general behavior of the foreground spillover with the CST beam.

\Cref{fig:cst_bleed-v_time} shows the spillover vs. hour angle for the CST beam with the GSM \cref{subfig:gsmcst} and sym \cref{subfig:symcst} models. The symmetric model more clearly demonstrates a pattern -- the spillover is generally higher for the shorter baseline, because total power is higher on the shorter baseline. When the galactic center is near the horizon, the baseline length becomes less important to the spillover and the curves come together. This is because the strongest foreground component, near the galactic center, is near to the horizon in delay space, so its power extends further beyond the horizon than when it is at a smaller delay.

\begin{figure}
    \centering
    \includegraphics[width=\columnwidth]{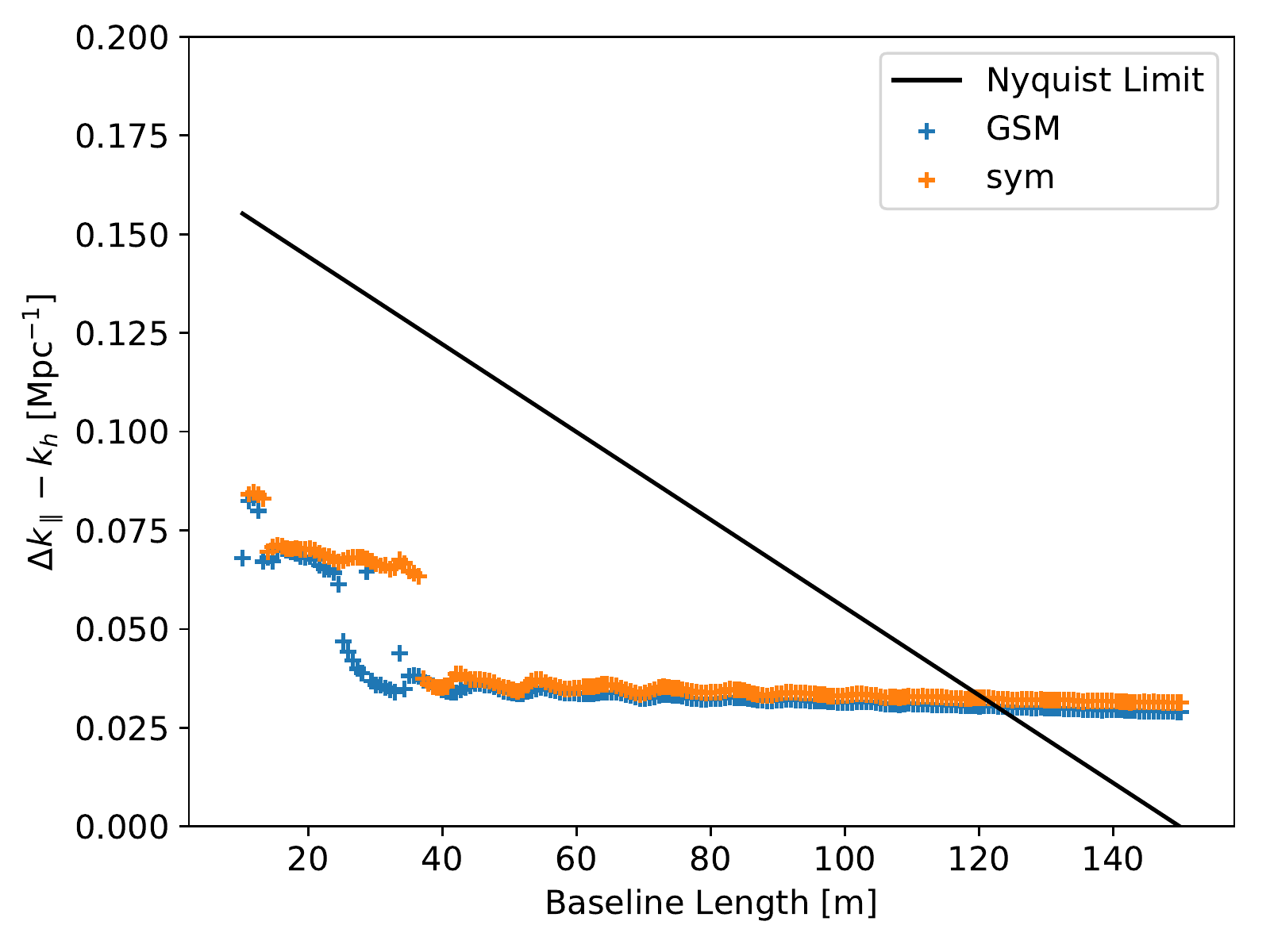}
    \caption{The spillover vs baseline length for the CST beam, for GSM (blue) and sym (orange) models. The black line indicates the Nyquist frequency cutoff of the UVBeam model. On especially long baselines, the spillover exceeds this limit, but continues in the general trend seen before. The jumps near small baseline length are due to the increase spectrum amplitude shifting some of the features at higher $k_\parallel$ above the threshold.}
    \label{fig:cst_bleed_v_bllen}
\end{figure}

\Cref{fig:cst_bleed_v_bllen} shows the spillover vs. baseline length for the GSM and sym models. The overall decrease with baseline length is apparent, as in \cref{fig:bleed_v_bllen}, but is much lower overall than for the Airy beam.

\section{Summary}

Smooth-spectrum foreground sources are expected to be well-behaved in Fourier space, which leaves a clear and well-defined window in which to seek an EoR detection. Spectral structure of the foregrounds, chromatic effects of the baseline and primary beam, and chosen bandpass and window function all contribute to the breakdown of this simple picture and the spread of power into the EoR window.

We carried out a series of simulations to characterize and measure this power spread for diffuse models in a variety of controlled cases. We used the Global Sky Model, and a symmetrized version of it with varied spectral steepness, in comparison to a fiducial EoR amplitude threshold. The analytic beams, antenna layouts, and frequency channelization and time step sizes are all chosen to resemble the design of HERA.

The choice of a spectral window function has the strongest effect on suppression of foreground power beyond the horizon; windowing along the frequency axis is required to reach the EoR level at all. Applying a window function effectively narrows the bandpass, which means that the main lobe of foreground power is made wider. The lowest $k_\parallel$ modes are therefore lost to foreground spillover. This trade-off between foreground suppression and spillover is necessary for a detection without substantial foreground removal.

For most of potential observing time, the most significant source of suprahorizon power appears to be the foreground pitchfork, which exists for primary beams with a response near the horizon. The pitchfork arises from the shortening of projected baselines near the horizon. As short baselines are more sensitive to diffuse structure, even faint monopole power near the horizon produces a strong response. This response can be observed for long baselines which are normally insensitive to diffuse power.

It is worth noting that the simulations used here do not take into account mutual coupling effects among antennas, as discussed in \cite{fagnoni_electrical_2019}. In a real array, the response of an antenna to sources near the horizon will be affected by the presence of the other antennas that interfere with the propagation of light. To take an extreme case, a bright source exactly in line with both antennas in a baseline will only be visible to the antenna nearer to it, since the farther antenna is in the near one's shadow. In this case, the correlation between the two antennas is zero. However, it may be that the effect of baseline projection is still significant enough to produce a pitchfork when emission is near the horizon but not in the shadow of other antennas. The full effects of mutual coupling are too sophisticated an effect to model with \textsc{healvis} at this time, and so we leave this to future work.

The Airy disk beam and CST simulated power beam both have a strong response near the horizon, and so have a constant foreground spillover due to this pitchfork for most times and baseline lengths. Avoiding LSTs when the galactic center is near the beam center, and choosing an appropriate constant buffer beyond the foreground wedge, appears to be sufficient to avoid diffuse foreground power, absent other systematic effects. It is difficult to draw a conclusion about foreground suppression at the level of the EoR for the CST beam, since those power spectra do not reach the EoR amplitude within the Nyquist limit of the CST simulations. What is apparent is that some degree of foreground \emph{subtraction} is necessary for an avoidance strategy to work.

\section{Acknowledgements}

This work was supported by NSF Grants 1613040 and 1636646, and NASA Grant 80NSSC18K0389. Computational support was provided by the Brown Center for Computation and Visualization (CCV). We thank the developers of \textsc{scipy}, \textsc{scikit-learn}, \textsc{matplotlib}, \textsc{astropy}, \textsc{healpy}, and \textsc{pyuvdata}.

%\afterpage{\clearpage}

\begin{appendices}
\crefalias{section}{appsec}
\section{Point source approximation}
\label{app:appendix_point_source}

The simulator used in this paper assumes that variations in the primary beam, brightness, and interferometric fringe are negligible on the scale of a single pixel, such that the contribution of each pixel can be treated as that of a point source at the pixel center. This appendix derives a rough bound on the error due to this \emph{point source approximation} (PSA) for the sky and beam models used in this paper. A forthcoming paper will explore the validity of this simulator against analytic solutions for diffuse sky models, and will delve into more detail on the limitations of the PSA.

We test the validity of our simulator here by performing a set of monopole simulations at different resolutions with an antenna layout featuring a broad range of baseline lengths. Our test array consists of 80 randomly-distributed antennas, forming baselines with lengths spanning 2 to 608~m, with the Airy beam and a bandpass of 100 to 120~MHz at 97 kHz channel width. The monopole sky makes for a useful test because it should minimize the power within the foreground wedge, which makes any erroneous simulated power stand out.

\begin{figure}
\includegraphics[width=\columnwidth]{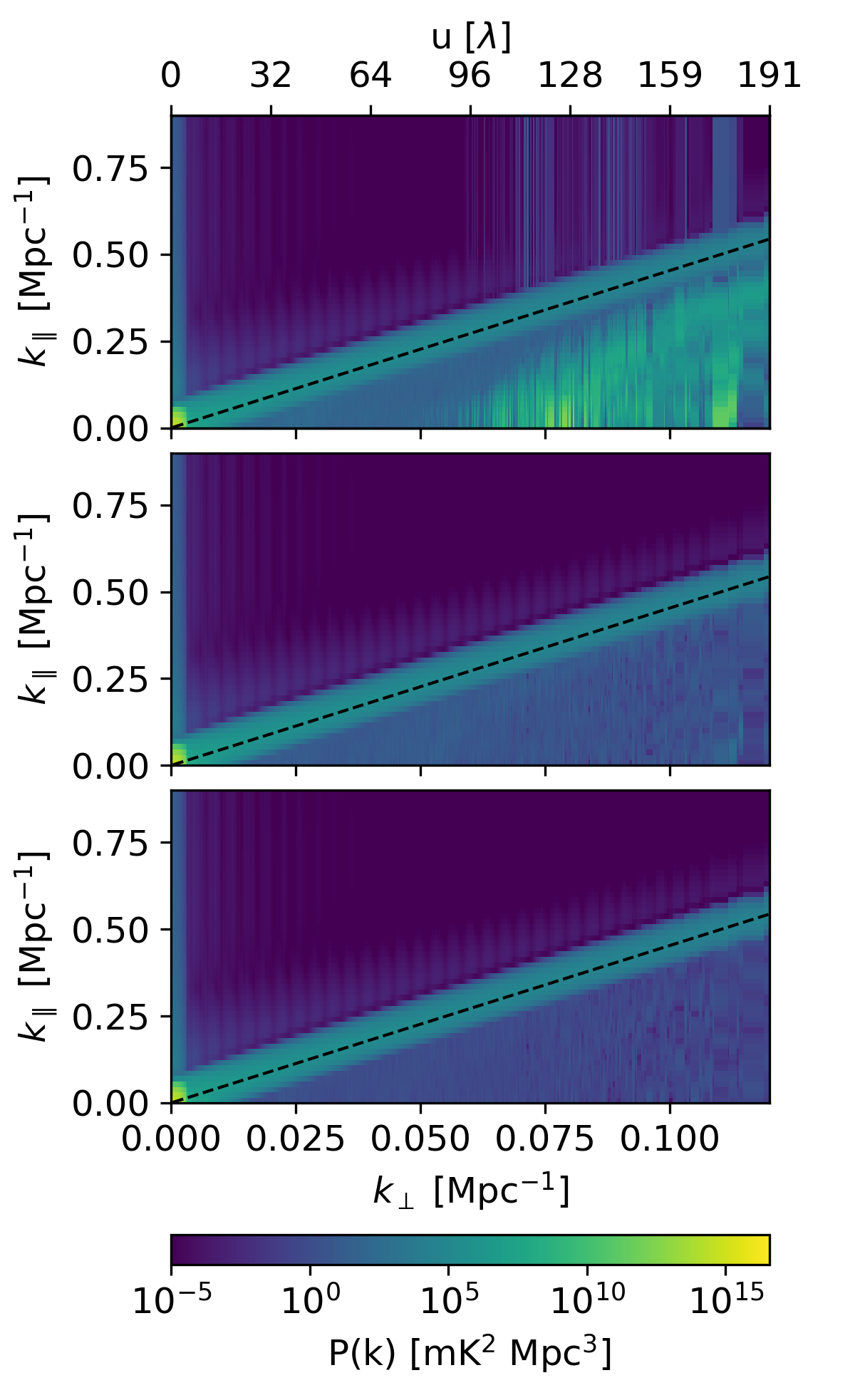}
\caption{Power spectra for a monopole signal on HEALPix maps with Nside 128, 256, and 512 (from top to bottom). The top axis shows the baseline length corresponding with the $k_\perp$ mode, and the dashed line marks the horizon. Excess power within the foreground wedge is erroneous for a monopole signal, and is clearly visibible at larger $k_\perp$ for the Nside 128 simulations. Clearly errors are reduced by switching from Nside 128 to 256, but further resolution improvement does not yield much noticeable improvement. Note that the longest baseline used for any results in this paper is at 50$\lambda$, for the spillover vs. baseline length tests. Further, the foreground pitchfork is not strongly influenced by the resolution of the map, as expected since it is a result of the projected baseline lengths being small near the horizon.}
\label{fig:restests}
\end{figure}

\Cref{fig:restests} shows the cylindrically-binned power spectra for these three simulations, for maps with Nside 128, 256, and 512 from top to bottom. Note that the longest baseline used in this paper, for the ``spillover vs. baseline length'' results, was only 50$\lambda$. The pitchfork feature is clearly visible in each plot around the horizon (marked with the dashed line). At Nside 128, there is excess power on long baselines (equivalently, at large $k_\perp$) which disappears at Nside 256. There is marginal improvement from raising the resolution again to Nside 512.

% The power in the pitchfork comes from integration over points where $|u'|$ is smallest, hence, the error \cref{eqn:psa_err} is small.
The pitchfork is largely independent of map resolution, which is further evidence that it is caused by the projection of baselines near the horizon. The error in the point source approximation is worse for longer baselines, but the power in the pitchfork comes from the shortening of baselines near the horizon. We note also the presence of faint vertical striping, and nonzero power within the wedge for all resolutions. The \emph{exact} visibility for a monopole sky signal with an Airy beam is given by the convolution of a circular disk (of diameter 14~m, in this case) with a sinc function of the baseline length. We suspect that the sidelobes of this sinc function are the cause of the stripes, as well as the broad spread of power at small $k_\perp$.

\end{appendices}

\bibliography{fg_bleed}
\bibliographystyle{mnras}

% Don't change these lines
\bsp	% typesetting comment
\label{lastpage}
\end{document}